\documentclass[iop]{emulateapj}
\usepackage{amsmath}
\usepackage{verbatim}
\usepackage{placeins}

\newcommand{\appropto}{\mathrel{\vcenter{
  \offinterlineskip\halign{\hfil$##$\cr
    \propto\cr\noalign{\kern2pt}\sim\cr\noalign{\kern-2pt}}}}}

\slugcomment{Draft version \today}

\shorttitle{Mass Evolution in the Perseus Molecular Cloud}
\shortauthors{Andersen et al.}

\begin{document}


\title{The Mass Evolution of Protostellar Disks and Envelopes in the Perseus Molecular Cloud}


\author{Bridget C. Andersen\altaffilmark{1,2}$\sp{\scriptscriptstyle\bigstar}$, Ian W. Stephens\altaffilmark{1}, Michael M. Dunham\altaffilmark{3,1}, Riwaj Pokhrel\altaffilmark{4}, Jes K. J{\o}rgensen\altaffilmark{5}, S{\o}ren Frimann\altaffilmark{5}, Dominique Segura-Cox\altaffilmark{6,7}, Philip C. Myers\altaffilmark{1}, Tyler L. Bourke\altaffilmark{8}, John J. Tobin\altaffilmark{9}, {\L}ukasz Tychoniec\altaffilmark{10,11}}
\email{$\sp{\scriptscriptstyle\bigstar}$E-mail: bca9vh@virginia.edu}


\altaffiltext{1}{Harvard-Smithsonian Center for Astrophysics, 60 Garden Street, MxS 78, Cambridge, MA 02138, USA}
\altaffiltext{2}{Department of Astronomy, University of Virginia, Charlottesville, VA 22904-4325, USA}
\altaffiltext{3}{Department of Physics, State University of New York at Fredonia, 280 Central Ave, Fredonia, NY 14063, USA}
\altaffiltext{4}{Department of Astronomy, University of Massachusetts, Maherst, MA 01003, USA}
\altaffiltext{5}{Center for Star and Planet Formation, Niels Bohr Institute and Natural History Museum of Denmark, University of Copenhagen, DK-1350 Copenhagen, Denmark}
\altaffiltext{6}{Department of Astronomy, University of Illinois, 1002 West Green Street, Urbana, IL 61801, USA}
\altaffiltext{7}{Max-Planck-Institute for Extraterrestrial Physics, Giessenbachstrasse 1, 85748 Garching, Germany}
\altaffiltext{8}{SKA Organization, Jodrell Bank Observatory, Lower Withington, Macclesfield, Cheshire SK11 9DL, UK}
\altaffiltext{9}{Department of Physics and Astronomy, University of Oklahoma, 440 W. Brooks St., Norman, OK 73019, USA}
\altaffiltext{10}{Leiden Observatory, Leiden University, P.O. Box 9513, NL-2300RA Leiden, The Netherlands}

\begin{abstract}
\indent In the standard picture for low-mass star formation, a dense molecular cloud undergoes gravitational collapse to form a protostellar system consisting of a new central star, a circumstellar disk, and a surrounding envelope of remaining material. The mass distribution of the system evolves as matter accretes from the large-scale envelope through the disk and onto the protostar. While this general picture is supported by simulations and indirect observational measurements, the specific timescales related to disk growth and envelope dissipation remain poorly constrained. In this paper we conduct a rigorous test of a method introduced by J{\o}rgensen et al. (2009) to obtain mass measurements of disks and envelopes around embedded protostars with observations that do not resolve the disk (resolution of $\sim$1000\,AU). Using unresolved data from the recent Mass Assembly of Stellar Systems and their Evolution with the SMA (MASSES) survey, we derive disk and envelope mass estimates for $59$ protostellar systems in the Perseus molecular cloud. We compare our results to independent disk mass measurements from the VLA Nascent Disk and Multiplicity (VANDAM) survey and find a strong linear correlation, suggesting that accurate disk masses can be measured from unresolved observations. Then, leveraging the size of the MASSES sample, we find no significant trend in protostellar mass distribution as a function of age, as approximated from bolometric temperatures. These results may indicate that the disk mass of a protostar is set near the onset of the Class 0 protostellar stage and remains roughly constant throughout the Class I protostellar stage.
\end{abstract}


\section{Introduction}
\indent Low-mass star formation is thought to begin when a dense molecular cloud undergoes gravitational collapse to form a protostellar system consisting of a new central star, a circumstellar disk, and a surrounding envelope of remaining material \citep[e.g.,][]{Shu1987,Allen2007,Dunham2014rev}. After the initial collapse, the system is in its earliest and most embedded (Class 0) stage. As the system evolves, matter is accreted from the large scale envelope through the disk and onto the protostar, eventually reaching the Class I stage when the protostar has accreted more than half its final mass \citep{Andre2000}. The disk plays an important role in this process, acting as an essential connection between the dissipating envelope and the growing star. \newline
\indent Disk formation occurs as a result of the conservation of angular momentum in the collapsing cloud. While this is expected to occur during the protostellar stage, the timescales related to disk growth and envelope dissipation are poorly constrained \citep{Dunham2014sim}. Theoretical models of protostellar disk formation allow for many possible outcomes. Non-magnetic hydrodynamic models of collapsing clouds indicate the formation of massive and extended disks in the early embedded stages of protostellar evolution \citep{Yorke1999,Vorobyov2009}. However, ideal magnetohydrodynamic models and simulations have found that strong magnetic braking may completely suppress disk formation in the early protostellar stage \citep[e.g.,][]{Allen2003,Mellon2008,Seifried2011}. Other magnetic models have explored physical processes that reduce the effects of magnetic braking and allow the formation of massive disks at later evolutionary stages \citep[e.g.,][]{Dapp2010,Machida2011}. For more details, \cite{Li2014} provides a comprehensive review of early disk formation and the effects of magnetic braking. \newline
\indent Direct mass measurements for disks around embedded protostars are difficult to determine because disk emission is typically entangled with emission from material in the surrounding envelope. Thus, one of the primary obstacles for studies of protostellar mass evolution involves devising accurate methods for separating disk and envelope emission. While data from telescopes like the Very Large Array (VLA) or the Atacama Large Millimeter Array (ALMA) can be used to resolve disk emission and obtain mass measurements, successfully proposing for large amounts of data from such highly subscribed instruments can be challenging. Since we wish to sample a large number of protostellar systems to derive robust statistical conclusions, it is more practical to search for ways to analyze disk masses in unresolved data, which are easier to obtain. \newline
\indent Recently, \citealt{Jorgensen2009} (hereafter J09) analyzed a sample of $10$ Class 0 and $10$ Class I protostellar systems in nearby ($\text{distance} \lesssim 300$\,pc) star-forming regions using unresolved $1.1$\,mm observations from the Submillimeter Array \citep[SMA;][]{Ho2004}. To accomplish this, J09 introduced a simple technique that uses the interferometric properties of SMA data to separate emission from the circumstellar disk and the surrounding envelope, allowing for the calculation of separate disk and envelope mass estimates for each protostar. To develop the framework for this method, J09 conducted radiative transfer simulations of collapsing power-law envelopes in nearby (distance $=125$\,pc) star-forming regions, generating mock $1.1$\,mm interferometric SMA and $850$\,$\mu$m single-dish James Clerk Maxwell Telescope (JCMT) data for each simulation. Using these mock data, J09 demonstrated that interferometric flux from the envelope is largely resolved out at baselines of $50$\,k$\lambda$ (corresponding to $\sim$500\,AU at $125$\,pc), contributing no more than $4\%$ of the corresponding single-dish flux. Therefore, by subtracting $4\%$ of the total flux found using JCMT/SCUBA legacy archive data \citep{DiFrancesco2008} from the $50$\,k$\lambda$ SMA flux, they were able to essentially isolate emission from the disk. The relative simplicity of the J09 method is easily scalable to larger (statistically significant) samples of protostellar disks. \newline
\indent Using the method described above, J09 found typical disk masses of $0.05$\,M$_{\odot}$ in both Class 0 and Class I protostellar systems. They found no evidence for significant disk mass growth between the Class 0 and Class I stages. They did, however, find that the disk-to-envelope mass ratio increased from Class 0 to Class I systems, indicating envelope dissipation over time. Based on the number of non-negligible Class 0 disks detected, J09 suggested that circumstellar disks are formed early in the Class 0 stage. \newline
\indent \cite{Dunham2014sim} later checked the validity of the J09 method with more detailed simulations combining non-magnetic hydrodynamical models of collapsing protostellar cores along with radiative transfer models. The increased complexity of their models in comparison to those used by J09 allowed them to test how the amount of envelope emission contamination in the interferometric flux changes as a function of dust opacity, optical depth, disk resolution, source distance, dust temperatures, and disk inclination. In particular, they determined that possible variations in the dust opacity laws due to grain growth could lead to order of magnitude overestimates in calculated disk masses at millimeter wavelengths. In addition, their simulations suggest that colder disk dust temperatures than the usual $30$\,K could lead to disk mass underestimates for Class I sources. Ultimately, \cite{Dunham2014sim} conclude that more observational checks are needed to truly test the validity of the J09 method. \newline
\indent Although large, the J09 sample was heterogeneous, composed of only the most well-known sources from different star forming regions. \cite{Enoch2011} speculated that the heterogeneous J09 sample could have led to environmental trends and selection biases in their results. To try to remove these biases, \cite{Enoch2011} used $1.3$\,mm data from the Combined Array for Research in Millimeter-Wave Astronomy (CARMA) to analyze a more homogeneous sample of $34$ Class 0 and Class I protostars in the Serpens molecular cloud. The selection of sources from the same molecular cloud ensured that distance uncertainties in mass estimates were reduced and that the star formation environments were similar (see \citealt{Dunham2014sim} for more information on how distance affects mass estimates). Using emission at $50$\,k$\lambda$ as an estimate of the disk flux without correcting for envelope contamination, \cite{Enoch2011} obtained a disk mass range of $0.04$\,M$_{\odot}$ to $1.7$\,M$_{\odot}$, with a mean of $0.2$\,M$_{\odot}$. Like J09, they did not detect significant mass growth between the Class 0 and Class I stages. They also found evidence of protostellar disks in $6$ of their $9$ Class 0 sources, consistent with the J09 suggestion of early (Class 0) disk formation. \newline
\indent Most recently, \citealt{Frimann2017} (hereafter F17) used $1.3$\,mm SMA and JCMT/SCUBA data along with the J09 method to analyze $24$ protostars in the Perseus molecular cloud. The SMA observations used in this study were obtained as part of the ``Mass Assembly of Stellar Systems and their Evolution with the SMA'' project (MASSES; Co-PIs: Michael M. Dunham and Ian W. Stephens; \citealt{Stephens2018}). They found a median disk mass of ${\sim}\,0.06$\,M$_{\odot}$. In addition, they performed a test of the J09 method by comparing the derived disk mass distribution of potential disk candidates in their sample (identified using results from previous resolved continuum observations) to the distribution of the whole sample. If the J09 method truly measures disk mass, then it might be expected that the disk candidates would have a higher mean disk mass than the non-candidates. However, F17 could not statistically reject the possibility that the disk candidates and non-candidates were drawn from the same distribution. This result indicates that the detection of high compact mass at long baselines does not necessarily demonstrate the presence of a disk, which brings into question the premise of the J09 method. \newline
\indent At the time of the F17 paper, the MASSES survey was still underway. Now that it has completed, the goal of the present study is to use the J09 method to obtain disk and envelope mass estimates of an even larger sample of $59$ sources from MASSES. We then compare our results to independent disk mass estimates from the resolved continuum observations in the VLA Nascent Disk and Multiplicity (VANDAM) survey (see \citealt{SeguraCox2018}, hereafter S18, and \citealt{Tychoniec2018}, hereafter T18, for methods and results). The purpose of this study is twofold. First, comparisons of our disk mass measurements to the VANDAM survey results provide a check on the validity of the J09 method. Second, with confirmation of the validity of the J09 method, the unprecedented size and uniformity of our sample places strong constraints on the timescales related to mass evolution in early protostellar systems. \newline
\indent This paper is organized as follows: Section~\ref{obsdata} introduces the SMA and JCMT/SCUBA-2 data and describes the protostellar systems in our sample; Section~\ref{analysis} describes the particular process used to obtain disk and envelope mass estimates; Section~\ref{discussion} contains a discussion of the results, in particular a comparison to the VANDAM disk mass measurements; and Section~\ref{summary} summarizes the main conclusions of the paper. \newline

\section{Observations and Data Reduction} \label{obsdata}
All targets observed in this paper are in the Perseus molecular cloud. The distance often used for Perseus is $235$\,pc \citep{Hirota2008}, which we adopt in this paper. While recent papers suggest the distance may more likely be $\sim$300\,pc \citep{Ortiz2018,Zucker2018}, we use 235\,pc for an easy comparison to other papers (i.e., S18 and T18).

\subsection{SMA Data}
\indent The bulk of the analysis in this paper is conducted on data from the SMA, an interferometer comprised of eight $6.1$\,m antennas located on Mauna Kea in Hawaii. In particular, the $1.3$\,mm continuum data are derived from the MASSES project, which employed the SMA to survey an unbiased and complete sample of all known embedded protostars in the Perseus molecular cloud, including $66$ sources identified with the \emph{Spitzer Space Telescope} by \cite{Enoch2009} and several additional young protostars (see \citealt{Stephens2018} for a full list of targets). In this study we analyze a subset of $59$ Class 0 and Class I sources from this sample (see Table~\ref{protoprops} for basic information about our sources). We only use the MASSES data that uses the Application Specific Integrated
Circuit (ASIC) correlator. \newline
\indent The continuum data presented in this paper were taken while the array was in its Subcompact configuration (covering baselines between ${\sim}\,5$\,k$\lambda$ and ${\sim}\,55$\,k$\lambda$), most typically with seven antennas in operation. These observations were taken in dual receiver mode with both the high frequency and low frequency receivers (centered on $850$\,$\mu$m or $356.72$\,GHz and $1.3$\,mm or $231.29$\,GHz, respectively) in operation. In this study, we only use the $1.3$\,mm data from the low frequency receiver. For each receiver, the correlator provided $2$\,GHz of bandwidth each to the lower and upper sidebands. \newline 
\indent The resulting visibility data were reduced and calibrated with the MIR software package using the standard procedure outlined in the MIR cookbook\footnote{See https://www.cfa.harvard.edu/$\sim$cqi/mircook.html}. The continuum data were then created by averaging eight correlator chunks of $64$ channels each, producing an effective frequency bandwidth of $1321$\,MHz. Finally, the data were cleaned and imaged across all baselines using the MIRIAD software package \citep{Sault1995}. See \cite{Lee2015} or \cite{Stephens2018} for more details about the MASSES data reduction process.

\subsubsection{Advantages of Using the MASSES Survey Data}
\indent The selection of our source sample from the MASSES survey combines several advantages only partially available to previous similar studies:
\begin{enumerate}
\item With $59$ analyzed sources, our study is the largest submillimeter project to date at ${\sim}\,1000$\,AU resolution. The number of sources allows us to derive robust statistical conclusions from our results.
\item Our sample includes a large portion of all known Class 0 and Class I sources in the Perseus molecular cloud. The completeness of our sample helps to reduce effects from selection bias, allowing us to obtain a balanced view of Class 0/I evolution. 
\item The restriction to a single molecular cloud helps us to reliably identify evolutionary trends distinct from environmental trends.
\end{enumerate}

\subsection{JCMT/SCUBA Data}
\indent This study also makes extensive use of $850$\,$\mu$m single-dish image data from the JCMT/SCUBA-2 Gould Belt Survey Data Release 1 (GBS DR1) \citep{Ward-Thompson2007,Holland2013}. The data from this release cover the brightest star-forming clumps in the Perseus molecular cloud, and include all of the sources in our sample. The GBS DR1 mosaics were created using the methodology described in \cite{Mairs2016} (and available to download through the Canadian Astronomical Data Centre\footnote{http://www.cadc-ccda.hia-iha.nrc-cnrc.gc.ca/en/jcmt/}; \citealt{Kirk2018}). The effective beam size of each of the mosaics is $14.6\arcsec$. \cite{Chen2016} presented the first analysis of the GBS Perseus observations using an earlier survey reduction (`Internal Release 1') which had larger pixel sizes and slightly poorer sensitivity to larger-scale structure than the version that we adopt here.

\section{Analysis} \label{analysis}
\indent To derive disk and envelope masses for each of the protostellar systems in our sample, we use a variation of the J09 method altered slightly to reflect that our SMA observations were taken at $1.3$\,mm instead of $1.1$\,mm. The resulting analysis is essentially equivalent to the procedure presented by F17. Fluxes extracted from $1.3$\,mm MASSES SMA data are used to measure compact emission while fluxes from JCMT/SCUBA-2 data are used to trace extended emission. Following the J09 procedure, we correct both the compact and extended emission for contamination from the envelope and disk, respectively. This produces isolated disk and envelope fluxes, which can then be converted to disk and envelope mass estimates. In the following section, we describe our complete method in detail.

\begin{figure}[!ht]
  	\centering
    \includegraphics[width=0.5\textwidth]{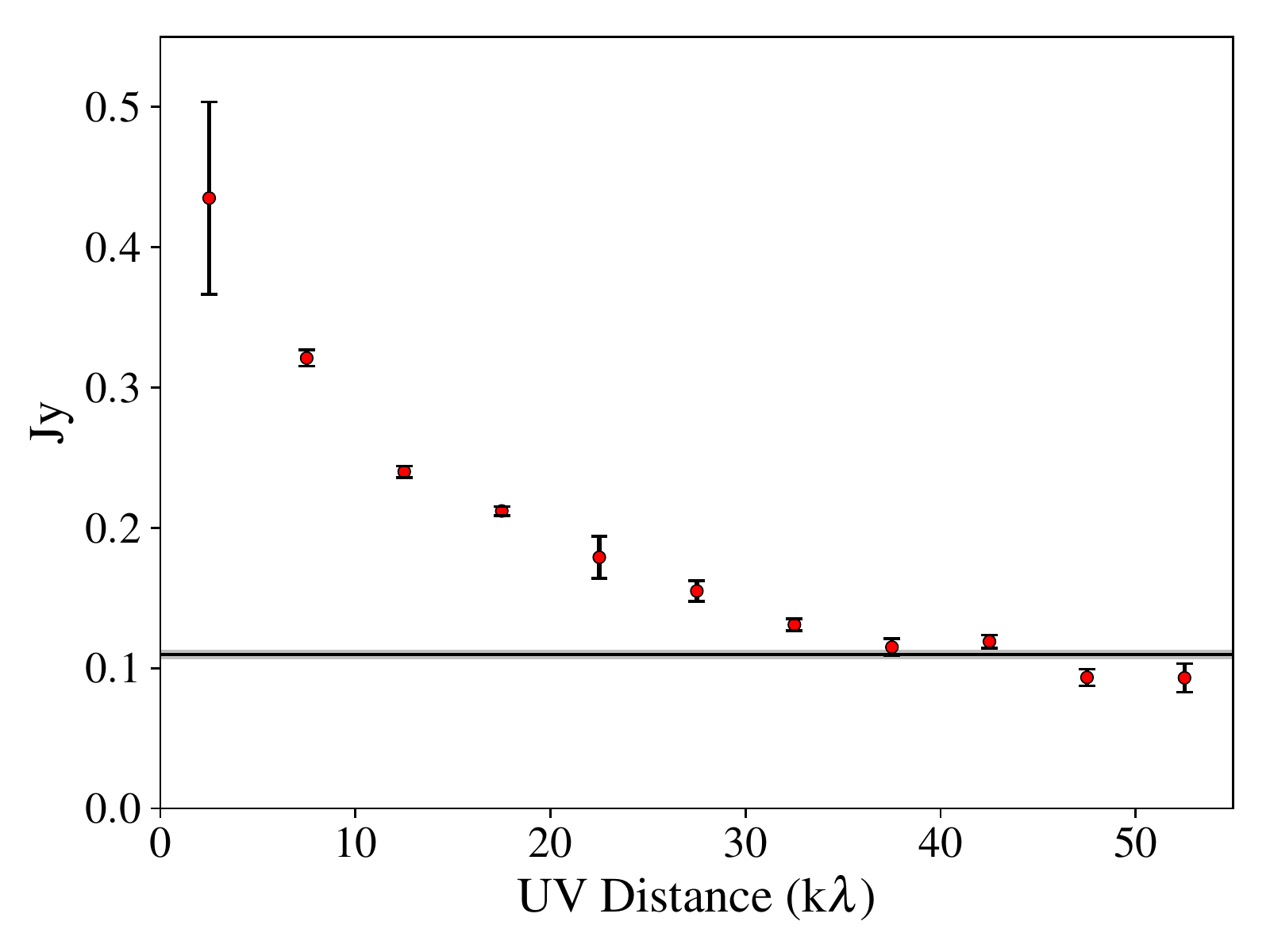}
    \caption{Amplitude of the flux density versus projected baseline for Per-emb-1. The black line shows the compact flux measured by fitting a \texttt{uvfit} point source to baselines ${>}40$\,k$\lambda$, and the gray area shows the error in the fit.} \label{Per1_uvamp}
\end{figure}

\subsection{1.3\,mm SMA Fluxes}
\indent As in F17, compact fluxes were extracted by fitting a point-source model to the SMA visibilities at baselines ${>}40$\,k$\lambda$ (corresponding to $\sim$1200\,AU at the distance of Perseus) using the \texttt{uvfit} MIRIAD routine. More specifically, the \texttt{cgcurs} routine was first used to produce a custom MIRIAD region file encompassing the compact source in the corresponding $1.3$\,mm continuum image across all baselines. This region file was then fed into the \texttt{imfit} routine which was used to simultaneously fit a point source and an elliptical Gaussian to the source in the image domain, generating initial estimates of the flux and position parameters of the point source. Finally, these estimates were passed as initial source parameter estimates (\texttt{spar} parameters) to the \texttt{uvfit} point source model, and the data were fit to determine the final compact flux at baselines ${>}40$\,k$\lambda$. The resulting point-source location derived from each fit was manually checked to ensure that it was coincident with emission from the compact source and not an unassociated local maximum. \newline
\indent This procedure was repeated for all compact sources in the MASSES survey images constituting our dataset. During our analysis, we found that $10$ of these images were of poor quality due to lack of uv-coverage or contained sources that were too faint to be reliably identified and fit (or sources that were simply not detected). These images were discarded (more specifically, the images centered on Per-emb-4, Per-emb-24, Per-emb-31, Per-emb-32, Per-emb-39, Per-emb-45, Per-emb-48, Per-emb-52, Per-emb-59, and L1448IRS2E). If there were multiple sources present in a single field of view, then they were all fit simultaneously with a single call to \texttt{uvfit}. In addition, many of the sources were imaged multiple times in different fields of view (e.g., Per-emb-12 is located just outside of the SMA's primary beam when it is centered on Per-emb-13). In these cases, the compact flux derived from the image with the source at the center of the field of view was used in further analysis since sensitivity is the highest toward the phase center. In the end, we were able to obtain flux measurements for $59$ unique sources. The sixth column in Table \ref{protoprops} displays the derived compact fluxes ($S_{{>}40\text{k}\lambda}$). \newline
\indent One of the primary challenges in using this method involves determining the ideal baseline lower limit for extracting only the compact flux component. In the uv-domain, a point source like an unresolved disk will approximately appear like a constant function spread across all baselines while emission from a surrounding protostellar envelope will appear as a Gaussian-like component peaking at ${\sim}\,0$\,k$\lambda$ (see Figure~\ref{Per1_uvamp} for an example of the flux density versus uv-distance for the source Per-emb-1). The lower baseline limit should be high enough to sufficiently resolve out the majority of the Gaussian envelope component while still retaining enough baselines to sample the constant point source component. J09 found that envelope emission is largely resolved out at $50$\,k$\lambda$. However, \cite{Dunham2014sim} also raised the concern that a baseline limit around $50$\,k$\lambda$ may start to resolve out emission from large nearby disks, leading to disk mass underestimates. \newline
\indent Taking into account all the factors described above, the ideal baseline lower bound is not entirely clear. To gain some sense of the uncertainty in the derived compact flux associated with the choice of baseline lower bound, we conducted several tests using the MIRIAD \texttt{uvfit} routine. In each test, we ran a \texttt{uvfit} point model fit with a baseline lower bound of $40$\,k$\lambda$ on uv-data containing a single solitary source (e.g., a single envelope system). We also completed a simultaneous fit of a point model and Gaussian model over all baselines. For this fit, the naive expectation is that the Gaussian models the extended envelope while the point models the remaining flux from the unresolved disk. However, in practice we found that the envelope emission was not always very well characterized by a Gaussian. Considering only sources that yielded reasonable Gaussian fits, we then took the relative difference between the ${>}40$\,k$\lambda$ point fit flux and the simultaneous point fit flux as an estimate our uncertainty. Completing this procedure for $5$ solitary sources, we found that our flux uncertainty ranged from ${\sim}\,2$ to $20\%$. Acknowledging this uncertainty, we ultimately chose a $40$\,k$\lambda$ baseline ($\sim$1200\,AU) lower limit to maintain consistency with F17.

\subsection{850\,$\mu$m JCMT/SCUBA-2 Fluxes}
\begin{figure}[!ht]
  	\centering
    \includegraphics[width=0.5\textwidth]{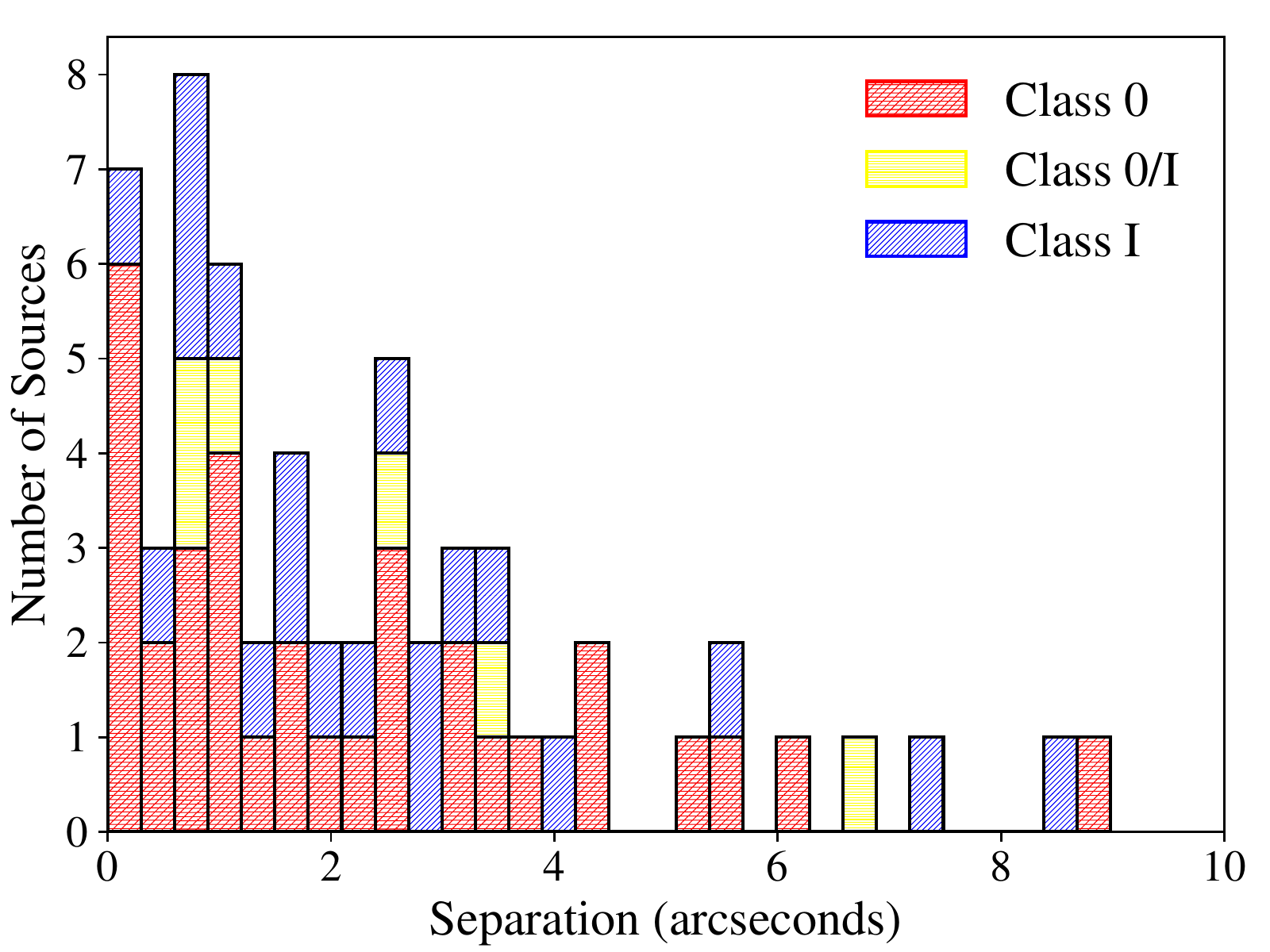}
    \caption{Stacked histogram of the separations between the \texttt{uvfit} location of the compact emission and \texttt{maxfit} location of the extended emission for each source, color-coded by source class. All of the separations are less than $14.6\arcsec$, the JCMT/SCUBA-2 beam size.} \label{pos_sep}
\end{figure}

\indent The extended flux of each protostellar source was extracted from the $850$\,$\mu$m single-dish JCMT/SCUBA-2 image data using the MIRIAD \texttt{maxfit} routine. Given a region defined using \texttt{cgcurs}, the \texttt{maxfit} routine finds the peak pixel within that region and then interpolates a maximum flux by fitting a parabola to a $3$ by $3$ grid centered on that pixel. In our analysis, we fit the $850$\,$\mu$m peak nearest to each compact source location determined from the previous \texttt{uvfit} results. We then took the resulting interpolated flux as the extended flux of the corresponding protostellar system. The seventh column of Table \ref{protoprops} displays the derived extended fluxes ($S_{14.6\arcsec}$). \newline
\indent Associating each $1.3$ mm compact source with a nearby $850$\,$\mu$m extended peak was not always trivial as some of the compact sources were not directly coincident with an extended peak. Figure \ref{pos_sep} shows a histogram of the separations between the \texttt{uvfit} location of the compact emission and \texttt{maxfit} location of the extended emission for each source. For $41$ out of the $59$ sources, the location of the compact emission was within $3\arcsec$ (approximately JCMT's pointing inaccuracy) of a peak in the $850$\,$\mu$m single-dish image. The remaining $18$ sources were spread from $3\arcsec$ to $9\arcsec$. The separation between these compact sources and their extended counterparts could be caused by a combination of multiple uncertainties including the $3\arcsec$ pointing inaccuracy of JCMT/SCUBA-2 \citep{DiFrancesco2008}, complex large scale envelope structure, and the limited JCMT/SCUBA-2 resolution of $14.6\arcsec$. Despite these possible sources of uncertainty, we were able to associate each compact source with an extended peak within the JCMT/SCUBA-2 beam size.

\subsection{Separating Disk and Envelope Flux} \label{sepdiskmassflux}
\indent While envelope emission is largely resolved out at baselines ${>}40$\,k$\lambda$, it may still contribute to the interferometric flux. Similarly, although the extracted $850$\,$\mu$m flux primarily traces the large-scale envelope, it also contains emission from the unresolved compact disk. To develop a framework for disentangling disk and envelope emission from measured interferometric and single-dish fluxes, J09 conducted radiative transfer simulations of nearby power law envelopes with variable inner cavity radii (and density profiles $\rho$~$\propto$~$r^{-1.5}$). From these simulations, they showed that the envelope contribution to the interferometric flux measured at 50\,k$\lambda$ at $1.1$\,mm is no more than $4\%$ of the extracted $850$\,$\mu$m single-dish envelope flux. \newline
\indent Based on this information, J09 proposed a system of two equations that could be solved to obtain estimates of the isolated disk and envelope fluxes. In this paper we use the same method, except, similar to F17, we adjust our equations to reflect the fact that we observed our interferometric fluxes at $1.3$\,mm rather than $1.1$\,mm as in J09. F17 estimated the envelope contamination factor, $c$, to be $2\%$ at $1.3$\,mm rather than the $4\%$ predicted by J09 at $1.1$\,mm, and we scale the disk flux by $(1.3/0.85)^{\alpha}$ when we subtract it from the single-dish emission to find the envelope flux. Here $\alpha$~$\approx$~$2.5$ is the spectral index of the dust continuum emission from disks at millimeter wavelengths, as \cite{Jorgensen2007} determined from the scaling between fluxes at $800$\,$\mu$m and $1.3$\,mm in emission at baselines ${>}40$\,k$\lambda$ for a sample of $8$ Class 0 sources. Taking these changes into account, we arrive at:
\begin{equation} \label{sepeqs}
\begin{aligned}
	S_{{>}40\text{k}\lambda} &= S_{\text{disk}} + c \cdot S_{\text{env}} \\
	S_{14.6\arcsec} &= S_{\text{disk}}\left(1.3/0.85\right)^{\alpha} + S_{\text{env}}
\end{aligned}
\end{equation}
in which $S_{{>}40\text{k}\lambda}$ is the measured $1.3$\,mm SMA interferometric flux at ${>}40$\,k$\lambda$, $S_{14.6\arcsec}$ is the measured $850$\,$\mu$m single-dish JCMT/SCUBA-2 flux, $S_{\text{disk}}$ is the unknown isolated disk flux, and $S_{\text{env}}$ is the unknown isolated envelope flux. For each of our $59$ protostellar systems, we solved this system of equations to estimate the isolated disk ($S_{\text{disk}}$) and envelope ($S_{\text{env}}$) fluxes. The resulting estimates are shown in the eighth and ninth columns of Table \ref{protoprops}. \newline
\indent Since the contamination factor $c$ is model dependent, we investigated how the results in this paper change for for values of $c$ between 0.01 and 0.05. At the extremities (i.e., $c=0.01$ and $c=0.05$), the estimated envelope and disk masses typically change by only $\sim$10\% and 30\%, respectively. For our disk mass comparisons with other studies (Section~\ref{sec:vandam}), the calculated correlations are unaffected and the slope of the line changes by less than 10\%.

\subsection{Calculating Masses} \label{obtainingmasses}
\indent The final step in our analysis involved deriving disk and envelope mass estimates. At $1.3$\,mm, emission from disk dust can be assumed to be optically thin. Under this assumption, the dust emission from a disk of constant temperature is proportional to the disk dust mass. Thus, adopting a typical gas-to-dust ratio of $R_{\text{gd}} \approx 100$, we can estimate the total disk mass using the following equation \citep{Hildebrand1983}
\begin{equation} \label{m_disk_eq}
	M_{\text{disk}} = R_{\text{gd}} \frac{d^2 S_{\text{disk}}}{\kappa_{\nu} B_{\nu}(T_{\text{dust}})},
\end{equation}
where $M_{\text{disk}}$ is the estimated disk mass; $d$ is the distance to the disk, which for our purposes we assume to be $235$\,pc for all sources \citep{Hirota2008}; $S_{\text{disk}}$ is the isolated disk flux; and $B_{\nu}(T_{\text{dust}})$ is the blackbody distribution flux at $1.3$\,mm for a dust temperature of $T_{\text{dust}}$. \newline
\indent The $\kappa_{\nu}$ in Equation~\ref{m_disk_eq} represents the dust opacity in the disk, which varies approximately with frequency according to the power law $\kappa_{\nu} \propto \nu^{\beta}$,
where $\beta$ is the opacity index. For optically thin emission in the Rayleigh-Jeans limit,
the spectral index of the disk, $\alpha$, is related to the dust opacity index as $\alpha$~$\approx$~$\beta + 2$ \citep{BeckwithSargent1991}. We adopt $\kappa_{1.3\,\text{mm}}$~=~$0.899$\,cm$^2$g$^{-1}$, which is the dust opacity for OH5 coagulated dust grains with thin ice mantles at $1.3$\,mm calculated by \citet{Ossenkopf1994}. We note that the opacities calculated by \citet{Ossenkopf1994} exhibit a dust opacity index of $\beta$~=~$1.8$, implying a spectral index of $\alpha$~=~$1.8 + 2$~=~$3.8$, which is inconsistent with $\alpha$ that we assumed in Equation~\ref{sepeqs}. However, to be consistent with \citet{Jorgensen2007,Jorgensen2009}, and considering the fact that dust grains in the disk may be larger than assumed in \citet{Ossenkopf1994}, we will use 
$\alpha$~=~$2.5$ ($\beta$~=~$0.5$) in Equation~\ref{sepeqs}. To test how this choice of spectral index affects our derived disk masses, we keep our value of $\kappa_{\nu}$ constant (i.e., $\kappa_{1.3\,\text{mm}}$~=~$0.899$\,cm$^2$g$^{-1}$) in Equation~\ref{m_disk_eq} and vary the value of $\alpha$ from $2.5$ to $3.8$ in Equation~\ref{sepeqs}. The average difference between our derived disk masses with $\alpha$~=~$2.5$ versus $\alpha$~=~$3.8$ is $4.6\%$. \newline
\indent Through detailed non-magnetic hydrodynamic and radiative transfer simulations of collapsing protostellar cores, \cite{Dunham2014sim} identified numerous sources of uncertainty present in Equation~\ref{m_disk_eq} including flux overestimation due to remaining flux contamination from the surrounding envelope, flux underestimation due to possible partial resolution of large nearby disks, changing dust opacity laws due to disk grain growth over time, and additional variations in optical depth, dust temperature, and disk inclination. Additional uncertainties arise from the assumed gas-to-dust ratio and constant assumed distance. The combined effects of all these elements lead to a total uncertainty of up to an order of magnitude in the resulting disk mass estimate. \newline
\indent Previous applications of the J09 method used a uniform disk dust temperature of $T_{\text{dust}}$~=~$30$\,K for all sources, regardless of protostellar Class (J09; \citealt{Enoch2011}; F17). However, \cite{Dunham2014sim} argue that since Class I disks tend to be larger and more optically thick, they are more shielded and thus cooler than Class 0 disks. Therefore, to mitigate some of the disk mass uncertainty, \cite{Dunham2014sim} suggest using a bimodal dust temperature of $30$\,K for Class 0 sources (including Class 0/I) and a cooler $15$\,K for Class I sources. By applying the J09 method, \cite{Dunham2014sim} show that they are better able to recover the true disk mass distribution of their simulated data using the two dust temperatures, rather than a constant temperature at $30$\,K for both Classes. We adopt this modification into our analysis. The resulting disk masses are displayed in the tenth column of Table \ref{protoprops}. \newline
\indent In some instances, the calculated disk masses were slightly negative or very low in value. We therefore constrained the minimum disk masses that are detectable given the SMA sensitivity for ${>}40$\,k$\lambda$ baselines. For each observation, we selected only baselines ${>}40$\,k$\lambda$ and used the MIRIAD \texttt{uvmodel} routine to subtract a point source model of each protostellar source from the calibrated uv-dataset using the previously obtained \texttt{uvfit} values as input parameters. We then produced a dirty image of this dataset with the MIRIAD routine \texttt{invert} and measured the rms of the image using the \texttt{imstat} routine. We then substituted this rms for the variable $S_{\text{disk}}$ in Equation~\ref{m_disk_eq} to estimate the mass sensitivity ($\sigma_{M_\text{disk}}$) for each observation. Sources with calculated disk masses that are less than $3\sigma_{M_\text{disk}}$ are considered not statistically significant. In the twelfth column of Table~\ref{protoprops}, we assign $3\sigma_{M_\text{disk}}$ as upper bounds to their disk masses. \newline
\indent While the assumption of isothermal optically thin dust is a good approximation for disk environments, large-scale protostellar envelopes exhibit a more complicated distribution of temperatures due to heating by internal protostars. Again using their radiative transfer models of nearby envelopes, J09 were able to determine the following empirical equation for more robust conversion between envelope flux and mass:
\begin{equation} \label{m_env_eq}
	\resizebox{0.42\textwidth}{!}{$M_{\text{env}} = 0.44\,\text{M}_{\odot} \left(\frac{L_{\text{bol}}}{1 L_{\odot}} \right)^{-0.36} \left(\frac{S_{\text{env}}}{1 \text{ Jy beam}^{-1}} \right)^{1.2} \left(\frac{d}{125 \text{ pc}}\right)^{1.2}$},
\end{equation}
where $M_{\text{env}}$ is the unknown envelope mass, $L_{\text{bol}}$ is the bolometric luminosity of the source, $S_{\text{env}}$ is the isolated envelope flux, and $d$ is the distance to the source, assumed to be $235$\,pc for all sources. Again, this estimate of envelope mass is associated with an order of magnitude uncertainty from the gas-to-dust ratio, dust opacity, and envelope density profile assumed by J09. We obtain envelope mass estimates for each of our sources using Equation~\ref{m_env_eq} and the bolometric luminosities derived from \cite{Tobin2016} and references therein. The resulting envelope masses are displayed in the eleventh column of Table \ref{protoprops}.\newline
 \newline

\begin{figure}[!ht]
  	\centering
    \includegraphics[width=0.5\textwidth]{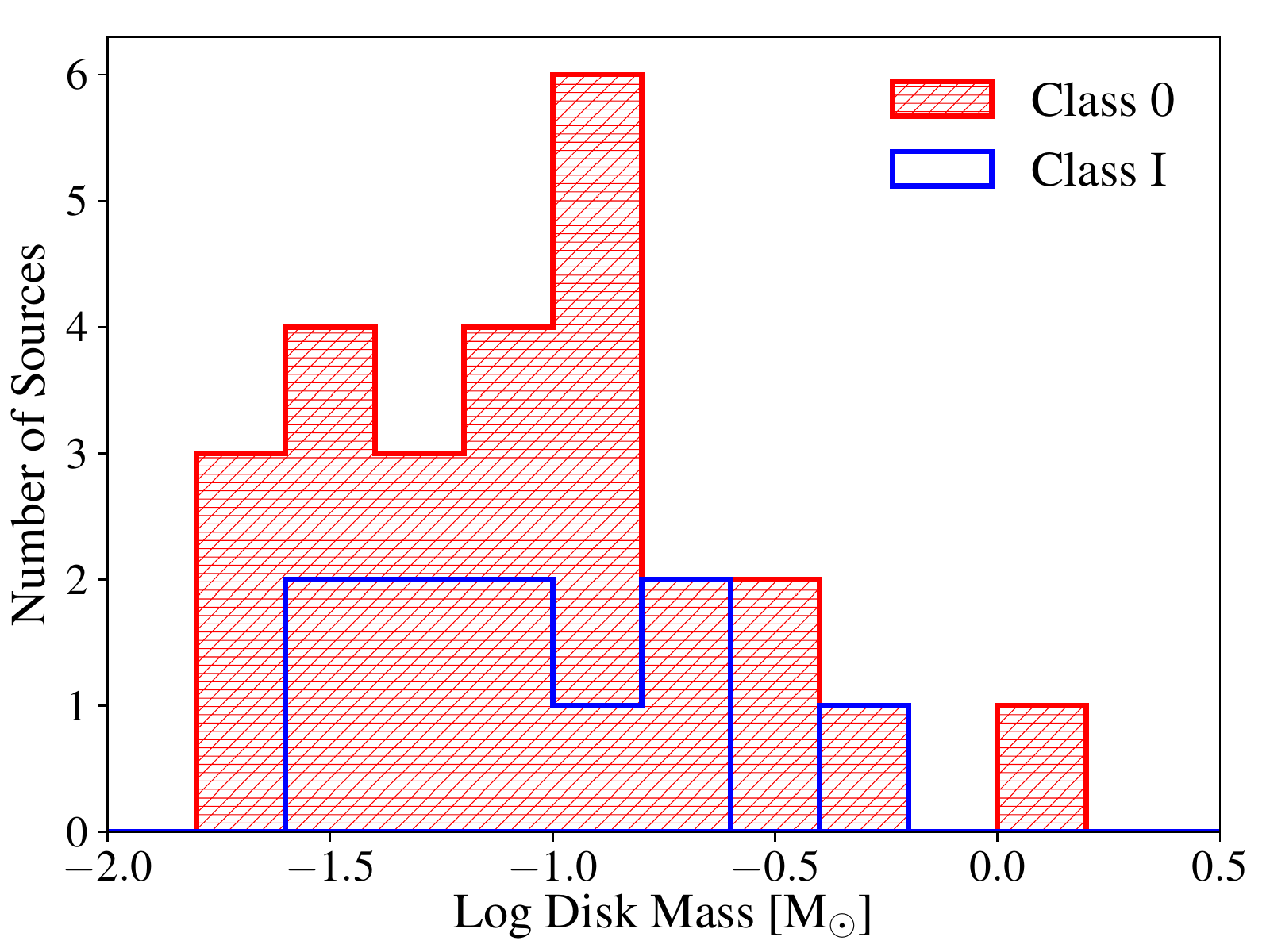}
    \caption{Histogram of the disk mass estimates in this paper, separated by protostellar classification} \label{disk_histo}
\end{figure}

\section{Results} \label{discussion}
\indent The measured fluxes and calculated masses for each source are shown in Table \ref{protoprops}, along with their corresponding protostellar classification and bolometric temperature obtained from \cite{Tobin2016} and references therein. A histogram of the disk masses (excluding those with $3\sigma_{M_\text{disk}}$ upper bounds) is shown in Figure~\ref{disk_histo}. In this figure, we show separate histograms for Class~0 and~I protostars, and we see no obvious difference in their distribution. \newline
\indent As a sanity check, we compare our corrected disk and envelope masses to those found by F17, who used a nearly identical method to analyze $24$ protostellar systems in the MASSES dataset. We find that our disk masses typically agree with the F17 results to within a median difference of ${\sim}\,14\%$; while our envelope masses are consistently smaller than theirs by a median of ${\sim}\,40\%$. We attribute the differences between the envelope masses as a result of two primary factors. First, F17 used older single-dish data from the JCMT/SCUBA legacy catalogue \citep{DiFrancesco2008}, while we used newer data from the JCMT/SCUBA-2 GBS DR1. The variations in calibration and quality between these two datasets may contribute an uncertainty to the extracted fluxes. Second, for the JCMT/SCUBA legacy catalogue used by F17, \cite{DiFrancesco2008} derived a beam size of about $23\arcsec$ due to smoothing during the data reduction. Given that Equation~\ref{sepeqs} uses intensity (i.e., is in units of Jy\,beam$^{-1}$), a larger beam adds more flux per beam since these protostars are embedded in extended structures. The difference in the beam sizes will therefore lead to overestimates in envelope flux, which would affect not only the derived envelope mass, but also the corrected disk mass. If we smooth the SCUBA-2 GBS DR1 data to the same beam size as the JCMT/SCUBA legacy survey, we find that our estimated disk and envelope masses agree with the F17 results, with a median difference of ${\sim}\,7\%$ and ${\sim}\,20\%$, respectively. \newline
\indent For four sources (Per-emb-6, Per-emb-28, Per-emb-54, and Per-emb-60), the corrected isolated disk fluxes are negative, indicating that we have overestimated the amount of envelope contamination in the ${>}40$\,k$\lambda$ interferometric flux. These particular sources could have larger inner cavities or shallower power-law envelopes than the $\rho$~$\propto$~$r^{-1.5}$ originally assumed by J09. Under these circumstances, the contamination fraction present in Equation~\ref{sepeqs} would be lower than our assumed value of $c$~=~$0.02$. \newline
\indent We also note that three sources (Per-emb-57, Per-emb-62, Per-emb-64) have negative isolated envelope fluxes. Since all of these sources are more evolved Class I sources with high bolometric temperatures (T$_{\text{bol}}$~$>$~$300$ K), it is possible that the envelopes surrounding these sources have almost completely dissipated (mass ${\sim}\,0$\,M$_{\odot}$). It is also possible that the spectral index (and the dust opacity index) for these sources is flatter than our assumed value of $\alpha$~$\approx$~$2.5$. A combination of these two factors could lead to an underestimate of the envelope mass when the disk flux $S_{\text{disk}}$ is subtracted from the single-dish flux $S_{14.6\arcsec}$ in Equation~\ref{sepeqs}.

\begin{center}
\begin{deluxetable*}{lcc@{}c@{}c@{}c@{}c@{}c@{}c@{}c@{}c@{}c@{}}
\tabletypesize{\tiny}
\tablewidth{0pt}
\tablecaption{Protostellar Source Data: Classes, Bolometric Temperatures, Continuum Fluxes, and Derived Masses \label{protoprops}}
\tablehead{
\colhead{Source}      & \colhead{R.A.$^{\text{a}}$} & \colhead{Dec.$^{\text{a}}$}     & \colhead{Class$^{\text{b}}$} & \colhead{T$_{\text{bol}}$ $^{\text{b}}$}      &
\colhead{$S_{{>}40\text{k}\lambda}$}          & \colhead{$S_{14.6\arcsec}$}  &
\colhead{$S_{\text{disk}}$$^{\text{c}}$}          & \colhead{$S_{\text{env}}$$^{\text{c}}$}    &
\colhead{$M_{\text{disk}}$$^{\text{c}}$}	&	\colhead{$M_{\text{env}}$$^{\text{c}}$} & \colhead{$M_{\text{disk,max}}$} \\
 & (J2000) & (J2000) & & \colhead{(K)} & \colhead{(mJy\,bm$^{-1}$)} & \colhead{(Jy\,bm$^{-1}$)} & \colhead{(mJy\,bm$^{-1}$)} & \colhead{(Jy\,bm$^{-1}$)} & \colhead{(M$_{\odot}$)} & \colhead{(M$_{\odot}$)} & \colhead{($3\sigma_{M_{\text{disk}}}$)}}
\startdata
Per-emb-1$^{\text{d,e}}$  & 03:43:56.77 & +31:00:49.87  & 0   & 27   & \phantom{$-$}0.110  & \phantom{$-$}1.51 & \phantom{$-$}0.085 & \phantom{$-$}1.26 & \phantom{$-$}0.061 & \phantom{$-$}1.15 & \nodata  \\
Per-emb-2$^{\text{d,e}}$  & 03:32:17.92 & +30:49:48.03   & 0   & 27   & \phantom{$-$}0.415 & \phantom{$-$}1.32 & \phantom{$-$}0.413 & \phantom{$-$}0.123 & \phantom{$-$}0.299 & \phantom{$-$}0.144 & \nodata \\
Per-emb-3$^{\text{e}}$   & 03:29:00.55 & +31:11:59.85   & 0   & 32   & \phantom{$-$}0.044 & \phantom{$-$}0.365 & \phantom{$-$}0.039 & \phantom{$-$}0.253 & \phantom{$-$}0.028 & \phantom{$-$}0.365 & \nodata \\
Per-emb-5$^{\text{d,e}}$& 03:31:20.93 & +30:45:30.33     & 0   & 32   & \phantom{$-$}0.208 & \phantom{$-$}0.622 & \phantom{$-$}0.207 & \phantom{$-$}0.023 & \phantom{$-$}0.150  & \phantom{$-$}0.023 & \nodata \\
Per-emb-6  &03:33:14.42 & +31:07:10.62  & 0   & 52   & \phantom{$-$}0.008 & \phantom{$-$}0.457 & $-$0.001               & \phantom{$-$}0.461 & $-$0.001               & \phantom{$-$}0.802 & 0.006 \\
Per-emb-8$^{\text{d,e}}$  & 03:44:43.98 & +32:01:34.97   & 0   & 43   & \phantom{$-$}0.107 & \phantom{$-$}0.642 & \phantom{$-$}0.099 & \phantom{$-$}0.355 & \phantom{$-$}0.072 & \phantom{$-$}0.283 & \nodata \\
Per-emb-9  & 03:29:51.88 & +31:39:05.52    & 0   & 36   & \phantom{$-$}0.013 & \phantom{$-$}0.420  & \phantom{$-$}0.004 & \phantom{$-$}0.408 & \phantom{$-$}0.003 & \phantom{$-$}0.552 & 0.008 \\
Per-emb-10 & 03:33:16.41 & +31:06:52.38    & 0   & 30   & \phantom{$-$}0.012 & \phantom{$-$}0.605 & \phantom{$-$}0.00   & \phantom{$-$}0.605 & \phantom{$-$}0.00  & \phantom{$-$}0.818 & 0.006 \\
Per-emb-11$^{\text{d,e}}$ & 03:43:57.06 & +32:03:04.67  & 0   & 30   & \phantom{$-$}0.190  & \phantom{$-$}1.42 & \phantom{$-$}0.172 & \phantom{$-$}0.922 & \phantom{$-$}0.125 & \phantom{$-$}0.897 & \nodata \\
Per-emb-12$^{\text{d,e}}$ & 03:29:10.49 & +31:13:31.37   & 0   & 29   & \phantom{$-$}1.74  & \phantom{$-$}8.96 & \phantom{$-$}1.66  & \phantom{$-$}4.16 & \phantom{$-$}1.20   & \phantom{$-$}2.32 & \nodata  \\
Per-emb-13$^{\text{e}}$ & 03:29:11.99 & +31:13:08.14    & 0   & 28   & \phantom{$-$}0.726 & \phantom{$-$}4.02  & \phantom{$-$}0.685 & \phantom{$-$}2.04 & \phantom{$-$}0.497 & \phantom{$-$}1.39 & \nodata  \\
Per-emb-14$^{\text{d,e}}$& 03:29:13.52 & +31:13:57.75    & 0   & 31   & \phantom{$-$}0.082 & \phantom{$-$}0.737 & \phantom{$-$}0.071 & \phantom{$-$}0.530  & \phantom{$-$}0.052 & \phantom{$-$}0.679 & \nodata \\
Per-emb-15  & 03:29:04.21 & +31:14:48.64   & 0   & 36   & \phantom{$-$}0.016 & \phantom{$-$}0.645 & \phantom{$-$}0.003 & \phantom{$-$}0.637 & \phantom{$-$}0.002 & \phantom{$-$}0.998 & 0.010 \\
Per-emb-16  & 03:43:51.00 & +32:03:23.86   & 0   & 39   & \phantom{$-$}0.011 & \phantom{$-$}0.467 & \phantom{$-$}0.002 & \phantom{$-$}0.462 & \phantom{$-$}0.001 & \phantom{$-$}0.724 & 0.007 \\
Per-emb-17$^{\text{e}}$  & 03:27:39.12 & +30:13:02.53   & 0   & 59   & \phantom{$-$}0.051 & \phantom{$-$}0.538 & \phantom{$-$}0.043 & \phantom{$-$}0.413 & \phantom{$-$}0.031 & \phantom{$-$}0.277 & \nodata\\
Per-emb-18$^{\text{d,e}}$ & 03:29:11.26 & +31:18:31.33 & 0   & 59   & \phantom{$-$}0.117 & \phantom{$-$}1.18 & \phantom{$-$}0.099 & \phantom{$-$}0.892 & \phantom{$-$}0.072 & \phantom{$-$}0.693 & \nodata \\
Per-emb-19  & 03:29:23.48 & +31:33:28.94    & 0/I & 60   & \phantom{$-$}0.011 & \phantom{$-$}0.275 & \phantom{$-$}0.006 & \phantom{$-$}0.257 & \phantom{$-$}0.004 & \phantom{$-$}0.418 & 0.007 \\
Per-emb-20  & 03:27:43.20 & +30:12:28.96   & 0/I & 65   & \phantom{$-$}0.012 & \phantom{$-$}0.381 & \phantom{$-$}0.004 & \phantom{$-$}0.369 & \phantom{$-$}0.003 & \phantom{$-$}0.368 & 0.007 \\
Per-emb-21$^{\text{e}}$  & 03:29:10.69 & +31:18:20.15   & 0   & 45   & \phantom{$-$}0.045 & \phantom{$-$}1.09 & \phantom{$-$}0.024 & \phantom{$-$}1.02 & \phantom{$-$}0.017 & \phantom{$-$}0.575 & \nodata\\
Per-emb-22$^{\text{e}}$ & 03:25:22.35 & +30:45:13.21   & 0   & 43   & \phantom{$-$}0.075 & \phantom{$-$}1.24 & \phantom{$-$}0.054 & \phantom{$-$}1.08 & \phantom{$-$}0.039 & \phantom{$-$}0.766 & \nodata \\
Per-emb-23 & 03:29:17.25 & +31:27:46.34    & 0   & 42   & \phantom{$-$}0.011 & \phantom{$-$}0.346 & \phantom{$-$}0.004 & \phantom{$-$}0.334 & \phantom{$-$}0.003 & \phantom{$-$}0.408 & 0.006 \\
Per-emb-25$^{\text{e}}$ & 03:26:37.49 & +30:15:27.90    & 0/I & 61   & \phantom{$-$}0.082 & \phantom{$-$}0.331 & \phantom{$-$}0.080  & \phantom{$-$}0.101 & \phantom{$-$}0.058 & \phantom{$-$}0.107 & \nodata \\
Per-emb-26$^{\text{e}}$ & 03:25:38.87 & +30:44:05.30   & 0   & 47   & \phantom{$-$}0.170  & \phantom{$-$}1.73 & \phantom{$-$}0.143 & \phantom{$-$}1.32 & \phantom{$-$}0.104 & \phantom{$-$}0.690 & \nodata  \\
Per-emb-27$^{\text{d,e}}$& 03:28:55.56 & +31:14:37.17   & 0/I & 69   & \phantom{$-$}0.247 & \phantom{$-$}2.70 & \phantom{$-$}0.205 & \phantom{$-$}2.11 & \phantom{$-$}0.148 & \phantom{$-$}0.822 & \nodata \\
Per-emb-28 & 03:43:50.99 & +32:03:07.97    & 0   & 45   & \phantom{$-$}0.007 & \phantom{$-$}0.429 & $-$0.001               & \phantom{$-$}0.432 & $-$0.001               & \phantom{$-$}0.554 & 0.007 \\
Per-emb-29$^{\text{e}}$ & 03:33:17.86 & +31:09:32.31    & 0   & 48   & \phantom{$-$}0.132 & \phantom{$-$}2.12  & \phantom{$-$}0.095 & \phantom{$-$}1.85 & \phantom{$-$}0.069 & \phantom{$-$}1.30  & \nodata \\
Per-emb-30$^{\text{d,e}}$& 03:33:27.30 & +31:07:10.19    & 0/I & 78   & \phantom{$-$}0.043 & \phantom{$-$}0.181 & \phantom{$-$}0.041 & \phantom{$-$}0.061 & \phantom{$-$}0.030  & \phantom{$-$}0.057 & \nodata \\
Per-emb-33$^{\text{d,e}}$ & 03:25:36.32 & +30:45:14.77   & 0   & 57   & \phantom{$-$}0.539 & \phantom{$-$}4.30 & \phantom{$-$}0.481 & \phantom{$-$}2.91 & \phantom{$-$}0.348 & \phantom{$-$}1.53 & \nodata  \\
Per-emb-35$^{\text{d,e}}$ & 03:28:37.12 & +31:13:31.24  & I   & 103  & \phantom{$-$}0.036 & \phantom{$-$}0.510  & \phantom{$-$}0.027 & \phantom{$-$}0.432 & \phantom{$-$}0.048 & \phantom{$-$}0.220 & \nodata  \\
Per-emb-36$^{\text{e}}$  & 03:28:57.36 & +31:14:15.61    & I   & 106  & \phantom{$-$}0.125 & \phantom{$-$}0.582 & \phantom{$-$}0.121 & \phantom{$-$}0.234 & \phantom{$-$}0.214 & \phantom{$-$}0.144 & \nodata \\
Per-emb-37  & 03:29:18.94 & +31:23:13.11   & 0   & 22   & \phantom{$-$}0.013 & \phantom{$-$}0.408 & \phantom{$-$}0.005 & \phantom{$-$}0.393 & \phantom{$-$}0.004 & \phantom{$-$}0.569 & 0.006 \\
Per-emb-40 & 03:33:16.65 & +31:07:54.81    & I   & 132  & \phantom{$-$}0.017 & \phantom{$-$}0.286 & \phantom{$-$}0.011 & \phantom{$-$}0.252 & \phantom{$-$}0.020  & \phantom{$-$}0.246 & 0.023 \\ 
Per-emb-44$^{\text{e}}$ & 03:29:03.72 & +31:16:03.30    & 0/I & 188  & \phantom{$-$}0.322 & \phantom{$-$}2.89 & \phantom{$-$}0.281 & \phantom{$-$}2.08 & \phantom{$-$}0.203 & \phantom{$-$}0.668 & \nodata \\
Per-emb-46 & 03:28:00.36 & +30:08:01.01  & I   & 221  & \phantom{$-$}0.008 & \phantom{$-$}0.071 & \phantom{$-$}0.007 & \phantom{$-$}0.050  & \phantom{$-$}0.013 & \phantom{$-$}0.088 & 0.019 \\
Per-emb-47 & 03:28:34.51 & +31:00:50.70    & I   & 230  & \phantom{$-$}0.009 & \phantom{$-$}0.043 & \phantom{$-$}0.009 & \phantom{$-$}0.018 & \phantom{$-$}0.016 & \phantom{$-$}0.019 & 0.017 \\
Per-emb-49 & 03:29:12.09 & +31:18:14.02 & I   & 239  & \phantom{$-$}0.015 & \phantom{$-$}0.396 & \phantom{$-$}0.008 & \phantom{$-$}0.374 & \phantom{$-$}0.014 & \phantom{$-$}0.407 & 0.036 \\
Per-emb-50$^{\text{d,e}}$ & 03:29:07.76 & +31:21:57.16   & I   & 128  & \phantom{$-$}0.098 & \phantom{$-$}0.553 & \phantom{$-$}0.093 & \phantom{$-$}0.286 & \phantom{$-$}0.164 & \phantom{$-$}0.104 & \nodata \\
Per-emb-51  & 03:28:34.52 & +31:07:05.47   & I   & 263  & \phantom{$-$}0.010  & \phantom{$-$}0.227 & \phantom{$-$}0.006 & \phantom{$-$}0.211 & \phantom{$-$}0.010  & \phantom{$-$}0.618 & 0.026 \\
Per-emb-53$^{\text{d,e}}$ & 03:47:41.58 & +32:51:43.75   & I   & 287  & \phantom{$-$}0.026 & \phantom{$-$}0.315 & \phantom{$-$}0.021 & \phantom{$-$}0.254 & \phantom{$-$}0.037 & \phantom{$-$}0.164 & \nodata \\
Per-emb-54 & 03:29:02.83 & +31:20:41.32 & I   & 131  & \phantom{$-$}0.014 & \phantom{$-$}1.11 & $-$0.009 & \phantom{$-$}1.14 & $-$0.017 & \phantom{$-$}0.466 & 0.028 \\
Per-emb-56  & 03:47:05.42 & +32:43:08.33   & I   & 312  & \phantom{$-$}0.015 & \phantom{$-$}0.098 & \phantom{$-$}0.014 & \phantom{$-$}0.058 & \phantom{$-$}0.025 & \phantom{$-$}0.081 & 0.031 \\
Per-emb-57$^{\text{e}}$ & 03:29:03.32 & +31:23:14.34    & I   & 313  & \phantom{$-$}0.026 & \phantom{$-$}0.043 & \phantom{$-$}0.027 & $-$0.034               & \phantom{$-$}0.048 & $-$0.092 & \nodata  \\
Per-emb-58 & 03:28:58.36 & +31:22:16.81    & I   & 322  & \phantom{$-$}0.012 & \phantom{$-$}0.202 & \phantom{$-$}0.008 & \phantom{$-$}0.178 & \phantom{$-$}0.015 & \phantom{$-$}0.237 & 0.015 \\
Per-emb-60 & 03:29:19.94 & +31:24:02.62 & I   & 363  & \phantom{$-$}0.004 & \phantom{$-$}0.246 & $-$0.001               & \phantom{$-$}0.249 & $-$0.002               & \phantom{$-$}0.444 & 0.014 \\
Per-emb-61  & 03:44:21.30 & +31:59:32.53   & I   & 371  & \phantom{$-$}0.010  & \phantom{$-$}0.113 & \phantom{$-$}0.008 & \phantom{$-$}0.091 & \phantom{$-$}0.014 & \phantom{$-$}0.170 & 0.026  \\
Per-emb-62$^{\text{d,e}}$& 03:44:12.97 & +32:01:35.29    & I   & 378  & \phantom{$-$}0.074 & \phantom{$-$}0.202 & \phantom{$-$}0.074 & $-$0.014               & \phantom{$-$}0.132 & $-$0.013 & \nodata              \\
Per-emb-63$^{\text{d,e}}$& 03:28:43.28 & +31:17:33.25    & I   & 436  & \phantom{$-$}0.018 & \phantom{$-$}0.146 & \phantom{$-$}0.016 & \phantom{$-$}0.100   & \phantom{$-$}0.028 & \phantom{$-$}0.089 & \nodata \\
Per-emb-64$^{\text{e}}$  & 03:33:12.85 & +31:21:23.95   & I   & 438  & \phantom{$-$}0.050  & \phantom{$-$}0.073 & \phantom{$-$}0.052 & $-$0.077               & \phantom{$-$}0.092 & $-$0.057 & \nodata               \\
Per-emb-65$^{\text{e}}$  & 03:28:56.30 & +31:22:27.69   & I   & 440  & \phantom{$-$}0.030  & \phantom{$-$}0.097 & \phantom{$-$}0.029 & \phantom{$-$}0.012 & \phantom{$-$}0.052 & \phantom{$-$}0.026 & \nodata \\
SVS13B$^{\text{d,e}}$  & 03:29:03.03 & +31:15:51.36      & 0   & 20   & \phantom{$-$}0.234 & \phantom{$-$}2.89 & \phantom{$-$}0.187 & \phantom{$-$}2.35 & \phantom{$-$}0.135 & \phantom{$-$}2.65 & \nodata  \\
SVS13C$^{\text{d,e}}$& 03:29:01.97 & +31:15:38.20        & 0   & 21   & \phantom{$-$}0.055 & \phantom{$-$}1.34 & \phantom{$-$}0.030  & \phantom{$-$}1.26 & \phantom{$-$}0.022 & \phantom{$-$}1.22 & \nodata  \\
B1-bN$^{\text{e}}$  & 03:33:21.20 & +31:07:43.93     & 0   & 14.7 & \phantom{$-$}0.152 & \phantom{$-$}1.36 & \phantom{$-$}0.132 & \phantom{$-$}0.976 & \phantom{$-$}0.096 & \phantom{$-$}1.66 & \nodata  \\
B1-bS$^{\text{e}}$  & 03:33:21.33 & +31:07:26.42 & 0   & 17.7 & \phantom{$-$}0.281 & \phantom{$-$}1.78 & \phantom{$-$}0.260  & \phantom{$-$}1.03 & \phantom{$-$}0.189 & \phantom{$-$}1.32 & \nodata  \\
L1448IRS3A$^{\text{e}}$ & 03:25:35.68 & +30:45:35.16 & I   & 47   & \phantom{$-$}0.132 & \phantom{$-$}4.30 & \phantom{$-$}0.049 & \phantom{$-$}4.16 & \phantom{$-$}0.087 & \phantom{$-$}2.11 & \nodata  \\
L1448NW & 03:25:36.46 & +30:45:21.43   & 0   & 22   & \phantom{$-$}0.055 & \phantom{$-$}1.76 & \phantom{$-$}0.021 & \phantom{$-$}1.70 & \phantom{$-$}0.015 & \phantom{$-$}1.69 & 0.016  \\
L1451-MMS$^{\text{e}}$ & 03:25:10.24 & +30:23:55.01 & 0   & 15   & \phantom{$-$}0.035 & \phantom{$-$}0.216 & \phantom{$-$}0.032 & \phantom{$-$}0.123 & \phantom{$-$}0.023 & \phantom{$-$}0.406 & \nodata \\
Per-Bolo-45   & 03:29:06.76 & +31:17:22.30    & 0   & 15   & \phantom{$-$}0.013 & \phantom{$-$}0.440 & \phantom{$-$}0.004 & \phantom{$-$}0.428 & \phantom{$-$}0.003 & \phantom{$-$}1.42 & 0.014 \\
Per-Bolo-58  & 03:29:25.42 & +31:28:14.21     & 0   & 15   & \phantom{$-$}0.009 & \phantom{$-$}0.271 & \phantom{$-$}0.003 & \phantom{$-$}0.261 & \phantom{$-$}0.002 & \phantom{$-$}0.865 & 0.007 \\
IRAS4B$^{\prime e}$  & 03:29:12.83 & +31:13:06.96      & 0   & 15   & \phantom{$-$}0.276 & \phantom{$-$}4.02  & \phantom{$-$}0.207 & \phantom{$-$}3.42  & \phantom{$-$}0.150  & \phantom{$-$}8.82 & \nodata
\enddata
\tablenotetext{a}{Location of the MASSES envelope. Most of these coordinates are from \citet{Pokhrel2018}. Accurate positions of the protostars themselves can be found in \citet{Tobin2016}.}
\tablenotetext{b}{Data obtained from \cite{Tobin2016} and references therein.}
\tablenotetext{c}{Negative values are consistent with 0, and are simply a result of the calculations in Equation~\ref{sepeqs}.}
\tablenotetext{d}{We compare our measured disk masses for each of these $18$ sources to those measured by \cite{SeguraCox2016} and S18 (see Section 4.1).}
\tablenotetext{e}{We compare our measured disk masses for each of these $36$ sources to those measured by T18 (see Section 4.1).}
\tablenotetext{ }{\textbf{Notes:} Uncertainties in the measured fluxes ($S_{{>}40\text{k}\lambda}$ and $S_{14.6\arcsec}$) are approximately $2$ to $20\%$. As discussed in Section~\ref{obtainingmasses}, uncertainties in the derived disk and envelope fluxes and masses approach an order of magnitude.}
\end{deluxetable*}
\end{center}
\twocolumngrid

\subsection{VANDAM Survey Comparison}\label{sec:vandam}
\begin{figure}[!ht]
  	\centering
    \includegraphics[width=0.5\textwidth]{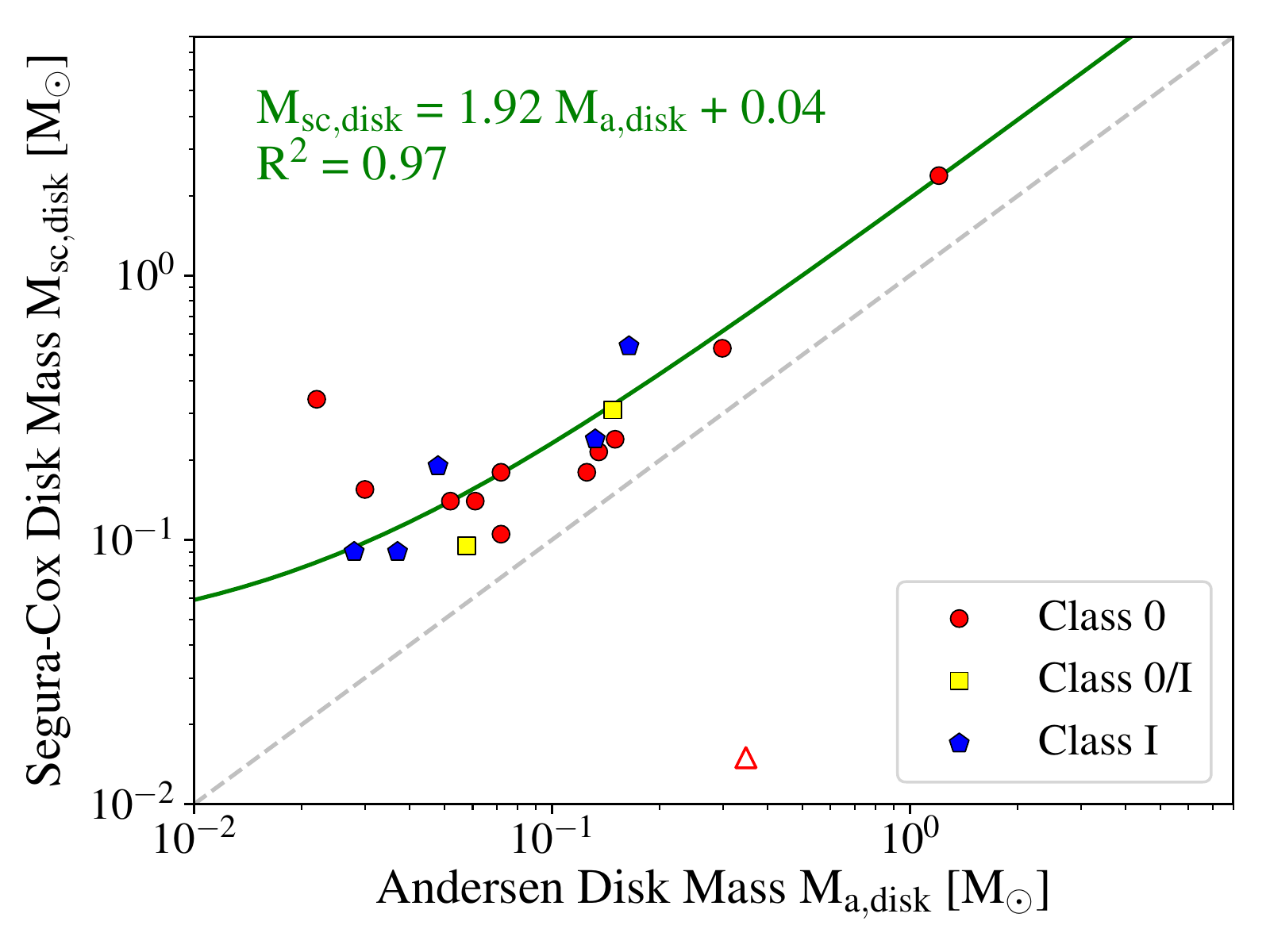}
    \caption{Comparison between our estimated disk masses and the corresponding disk masses measured using resolved continuum data from the VANDAM survey (see \citealt{SeguraCox2016} and S18). Class 0 sources are shown as red circles, Class 0/I sources are shown as yellow squares, and Class I sources are shown as blue pentagons \citep[as determined by][]{Tobin2016}. The red triangle represents the outlier data point corresponding to Per-emb-33. The grey dashed line represents the points where $M_{\text{sc,disk}} = M_{\text{a,disk}}$ while the green line shows a least-squares linear fit of the data excluding Per-emb-33. The resulting linear equation and coefficient of determination are labeled. The strong correlation suggests that the J09 method is valid for statistical studies of protostellar mass evolution.} \label{sc_compare}
\end{figure}

\begin{figure}[!ht]
  	\centering
    \includegraphics[width=0.5\textwidth]{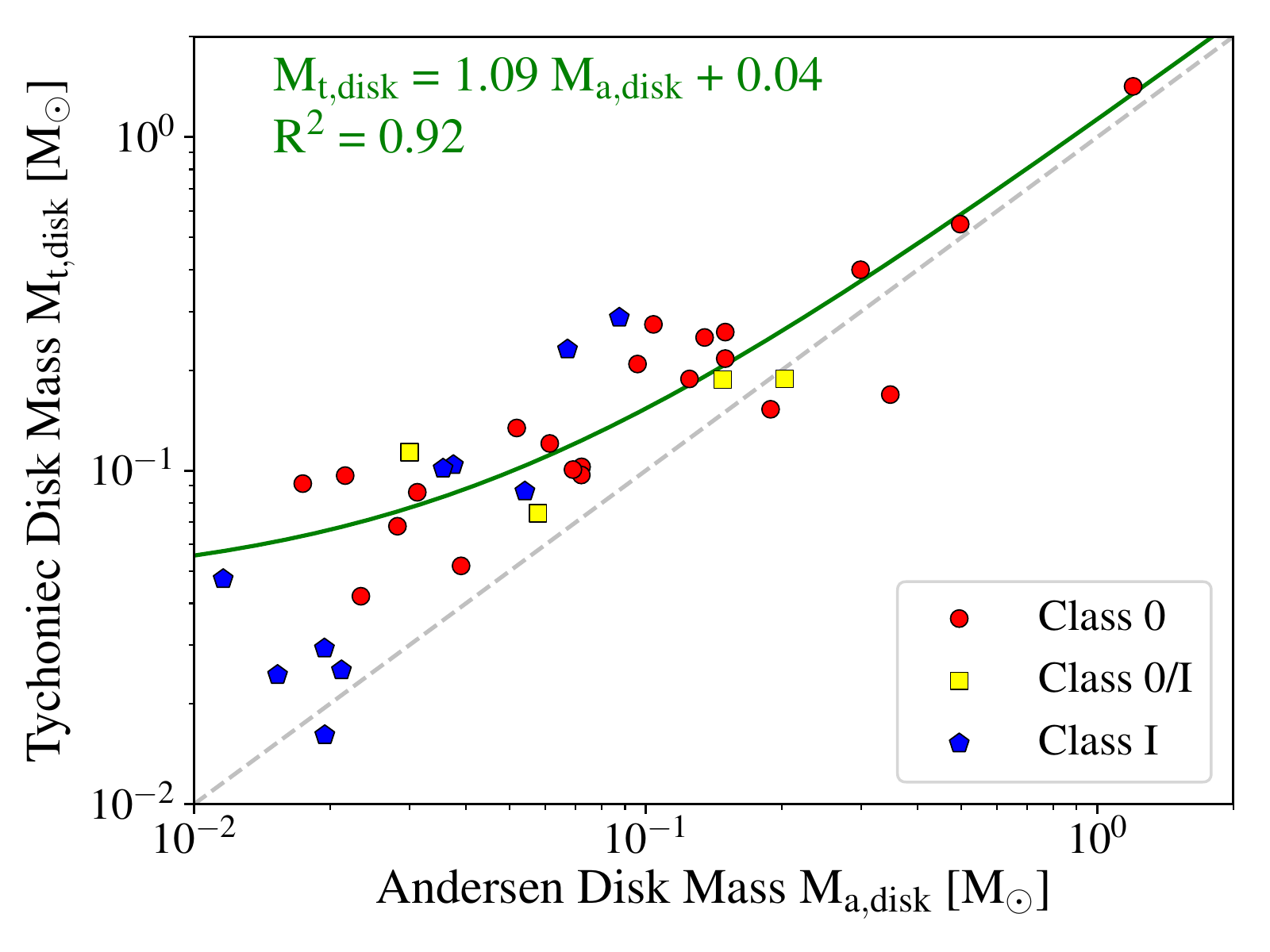}
    \caption{Comparison between our estimated disk masses and the corresponding disk masses measured using free-free corrected 8\,mm fluxes from the VANDAM survey (see T18). Colors of the points correspond to protostellar classification (see Figure~\ref{sc_compare}). The grey dashed line represents the points where $M_{\text{t,disk}}$~$=$~$M_{\text{a,disk}}$ while the green line shows a least-squares linear fit of the data. The resulting linear equation and coefficient of determination are labeled.} \label{tt_compare}
\end{figure}

\indent As a test of the validity of the J09 method, we compare our results to independent disk masses estimated from the VANDAM survey by S18 (see \citealt{SeguraCox2016} for initial results). The VANDAM survey constitutes a resolved VLA continuum ($12$\,AU resolution) survey toward all of the identified protostellar systems in the Perseus molecular cloud, including our sample \citep{Tobin2016}. Using data from this survey, S18 fit disk surface brightness profiles to resolved $8$\,mm continuum observations to identify disk candidates. For each candidate, they also derived a flux measurement and used Equation~\ref{m_disk_eq} to obtain a disk mass estimate (scaling $\kappa_{1.3\,\text{mm}}$~$=$~$0.899$\,cm$^2$g$^{-1}$ assuming a dust opacity index $\beta$~$=$~$1$). Since the disk temperature is uncertain, S18 derived a range of possible disk masses for each source, rather than a single value. Specifically, they varied the dust temperature, with $T_{\text{dust}}$~$=$~$20$\,K determining the upper mass limit and $T_{\text{dust}}$~$=$~$40$\,K determining the lower mass limit. \newline
\indent \cite{SeguraCox2016} identified and calculated disk mass ranges for $7$ disk candidate protostellar systems. More recently, S18 completed the same analysis for the rest of the sources in the VANDAM survey, identifying an additional $11$ disk candidates. Each of these $18$ sources are labeled by a superscript in the first column of Table 1. To properly compare our results for these sources, we extrapolated the VANDAM mass limits to obtain mass estimates at our assumed $T_{\text{dust}}$~$=$~$30$\,K and $15$\,K for Class 0 (including Class 0/I) and Class I sources, respectively. Since the blackbody distribution varies almost linearly with dust temperature ($M_{\text{disk}}$~$\propto$~$B_{\nu}(T_{\text{dust}})^{-1}$~$\appropto$~$T_{\text{dust}}^{-1}$), we simply take the average of the VANDAM mass limits to obtain estimates for Class 0 sources at an assumed temperature of $30$\,K. For Class I sources, we multiply this average by two to get the estimates at an assumed temperature of $15$\,K. \newline
\indent Figure \ref{sc_compare} shows a plot of our calculated disk masses compared to the corresponding disk masses extrapolated from the  \cite{SeguraCox2016} and S18 results. The red triangle indicates an outlier point corresponding to Per-emb-33. Per-emb-33 is a triple system of protostars embedded in a much larger disk \citep{Tobin2016b}. Since S18 only considered compact emission for a single protostar within this system, their estimated disk mass for the triple system is very underestimated. Moreover, the extended structure of Per-emb-33 has a low surface brightness and was not well-recovered by the VANDAM observations. Excluding Per-emb-33, we complete a simple linear least-squares fit of the rest of the data and find a strong linear correlation with a coefficient of determination of $R^2$~$=$~$0.97$. The resulting line of best fit is shown in green in Figure \ref{sc_compare}. While our disk masses tend to be ${\sim}\,2$ times smaller than the S18 estimates, the strength of the linear correlation between our results suggests that the J09 method is valid for estimating protostellar disk masses. The factor of two discrepancy between the disk masses may have a variety of causes, such as the dust opacity assumed by S18 ($\beta$~=~1), the density profile assumed by J09 ($\rho$~$\propto$~$r^{1.5}$), the contamination factor assumed by F17 ($c$~=~0.02), and the flux measurement uncertainty for the interferometers ($\sim$10\% for the VLA and $\sim$15\% for the SMA). \newline
\indent Examining our estimated disk masses for protostellar systems not identified as disk candidates by \cite{SeguraCox2016} and S18, we find $7$ sources that have masses greater than $0.1$\,M$_{\odot}$ (Per-emb-13, Per-emb-26, Per-emb-36, Per-emb-44, Per-emb-50, B1-bS, and IRAS4B$^{\prime}$). In particular, our estimated disk mass for Per-emb-13 (also known as NGC~1333 IRAS~4B) is ${\sim}\,0.5$\,M$_{\odot}$, which is higher than the majority of the disk candidate masses determined by S18. \newline
\indent S18 found the resolved 8\,mm emission from Per-emb-13 to be consistent with an envelope-only profile, and was therefore not identified as a disk candidate based on the VLA data. Because we find such a high estimated disk mass in the SMA data, the fact that Per-emb-13 is not a VANDAM disk candidate even though it is resolved at 8\,mm indicates some mechanism prohibits the detection of the disk at longer wavelengths. As one explanation, it is possible that the large dust grains in Per-emb-13 have condensed toward the center of the disk \citep{Weidenschilling1977,Birnstiel2010}. Since the VANDAM $8$\,mm continuum emission may be dominated by the largest grains, the disk component may be unresolved at $8$\,mm (while the envelope component is resolved), while in fact, the disk would look substantially larger at 1.3\,mm. The disk could also have complex, non-axisymmetric structure that made disk fitting invalid in S18. \newline
\indent It may also be possible that Per-emb-13 has a disk smaller than the ${\sim}\,0.05\arcsec$ ($12$\,AU) beam in the VANDAM survey, though a massive reservoir of large grains would have been detected as a point source component at 8\,mm, which is not seen in the VANDAM data. There is also a chance that no actual disk exists and our mass estimate purely reflects another compact feature. However, given that our disk estimates for $17$ sources correlate very strongly with those estimated by S18, it is likely that a disk exists. We note that S18 only model possible disk candidates at 8\,mm, and they do not rule out well-formed disks that may be detected at other wavelengths. Indeed, \citet{Persson2016} provides a possible disk model for Per-emb-13 based on observations at 1.5\,mm. \newline
\indent T18 also used VANDAM data to approximate disk masses. VANDAM observed each source in Perseus at 4 different wavelengths, with two in Ka-band at (8\,mm and 1\,cm) and two in C-band (4.1\,cm and 6.4\,cm). T18 used the C-band data to approximate the free-free and any non-thermal (i.e., synchrotron) component and subtracted that emission from the Ka-band at 8\,mm. They then use Equation~\ref{m_disk_eq} to calculate disk masses, using the same $\kappa_\nu$ is as \citealt{SeguraCox2016} and S18. This method does have some uncertainties associated with it. For example, they assume the free-free and any non-thermal emission are not variable (C-band and Ka-band observations were taken at different epochs), and then only perform an approximate correction when fitting fails or is unreliable due to variability. Moreover, they may only be sensitive to the largest grains in the disk. They also assume that the temperature scales with luminosity to the 0.25 power. While the method is expected to provide less accurate disk masses than the studies by Segura-Cox, they are capable of approximating masses of the unresolved sources, allowing for a much larger sample size. \newline
\indent In Figure~\ref{tt_compare}, we show the comparison between our estimated disks masses and those by T18. We plot only sources for which both T18 and we found significant disk masses (each of these 36 sources are labeled by a superscript in the first column of Table 1). Solely for the purposes of comparison, we have adjusted our disk mass estimates in this plot to assume a dust temperature of $30$\,K for both Class~0 and I sources. The correlation is strong ($R^2$~$=$~0.92) like the relationship found in the Segura-Cox comparison ($R^2$~$=$~0.97). Nevertheless, the slope of this line is 1 (i.e., there is no factor of two discrepancy as found in the Segura-Cox comparison), indicating that the masses measured by the two methods are strongly related.\newline
\indent In both this paper and T18, there were several protostars for which approximate disk masses could not be estimated. We find that every protostar without a disk mass measurement in T18 also did not have a disk mass measurement in this paper (i.e., at best we could only provide mass upper bounds). However, there were 14 sources for which we could only provide a disk mass upper bound, but T18 was able to estimate a mass. In particular, we find that for Per-emb-6, Per-emb-10, Per-emb-19, and Per-emb-37, the T18 disk masses were over a factor of 5 higher than our upper bound estimates. This discrepancy suggests that in at least some cases, our upper bound mass estimates could be inaccurate.

\begin{figure}[!ht]
  	\centering
    \includegraphics[width=0.5\textwidth]{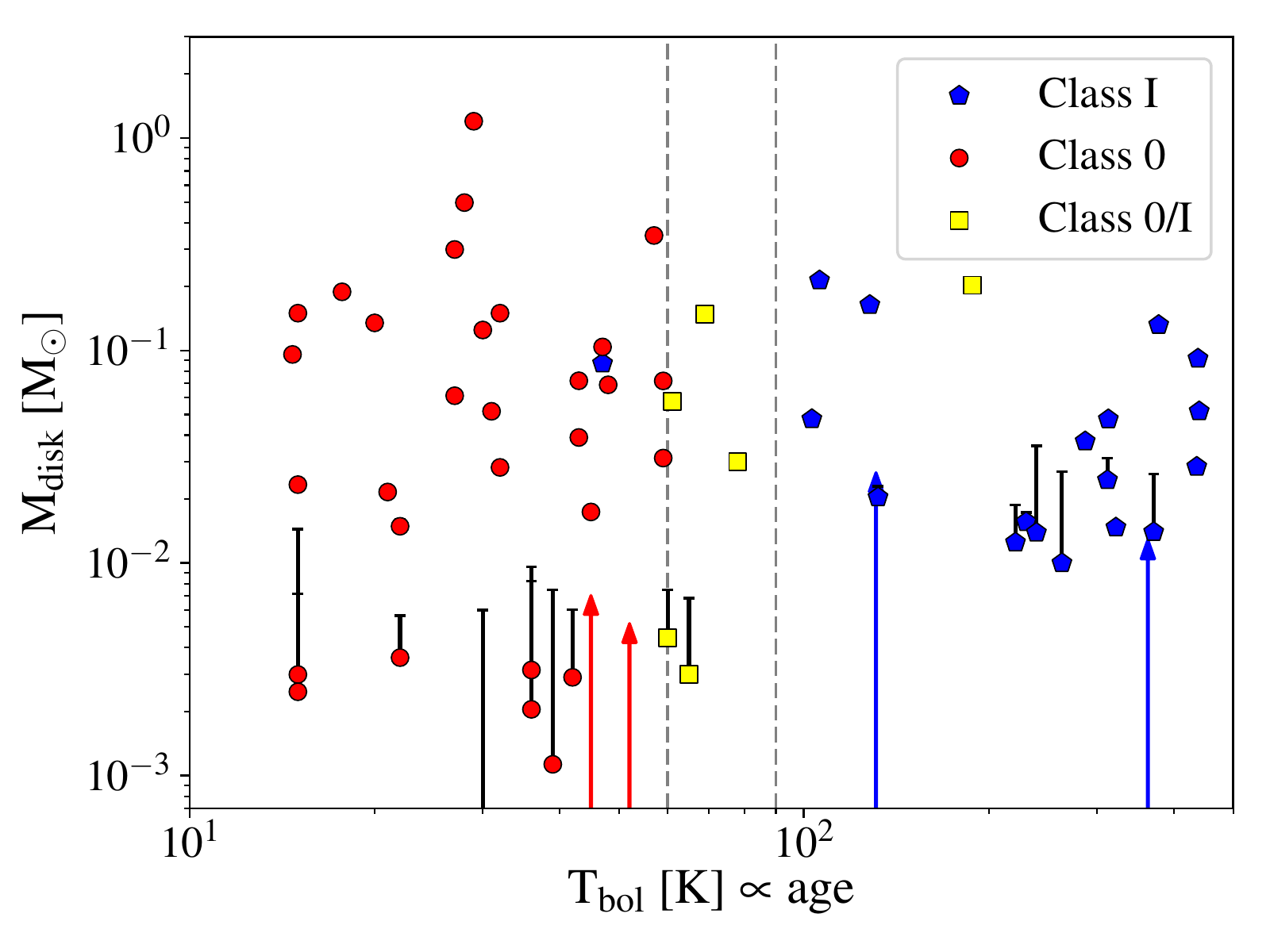}
    \caption{Estimated disk masses as a function of bolometric temperature, with the colors of the points corresponding to protostellar classification (see Figure~\ref{sc_compare}). The dashed gray lines separate Class 0, Class 0/I, and Class I sources based solely on bolometric temperature cutoffs \citep{Tobin2016}. Sources with upper error bars have calculated masses that are not statistically significant. The upper error bar shows the upper limit of the disk mass based on the sensitivity of the SMA observation (see Section~\ref{obtainingmasses}). Note that the error bar extending from the bottom of the plot (without a visible corresponding point) denotes the upper bound for Per-emb-10, which had a measured disk mass of approximately ${\sim}\,0$\,M$_{\odot}$. The arrows correspond to the upper limit for sources with negative estimated disk masses. The colors of the arrows are associated with their protostellar classification.} \label{m_disk_Tbol}
\end{figure}

\subsection{Resulting Disk and Envelope Masses}
\begin{figure}[!ht]
  	\centering
    \includegraphics[width=0.5\textwidth]{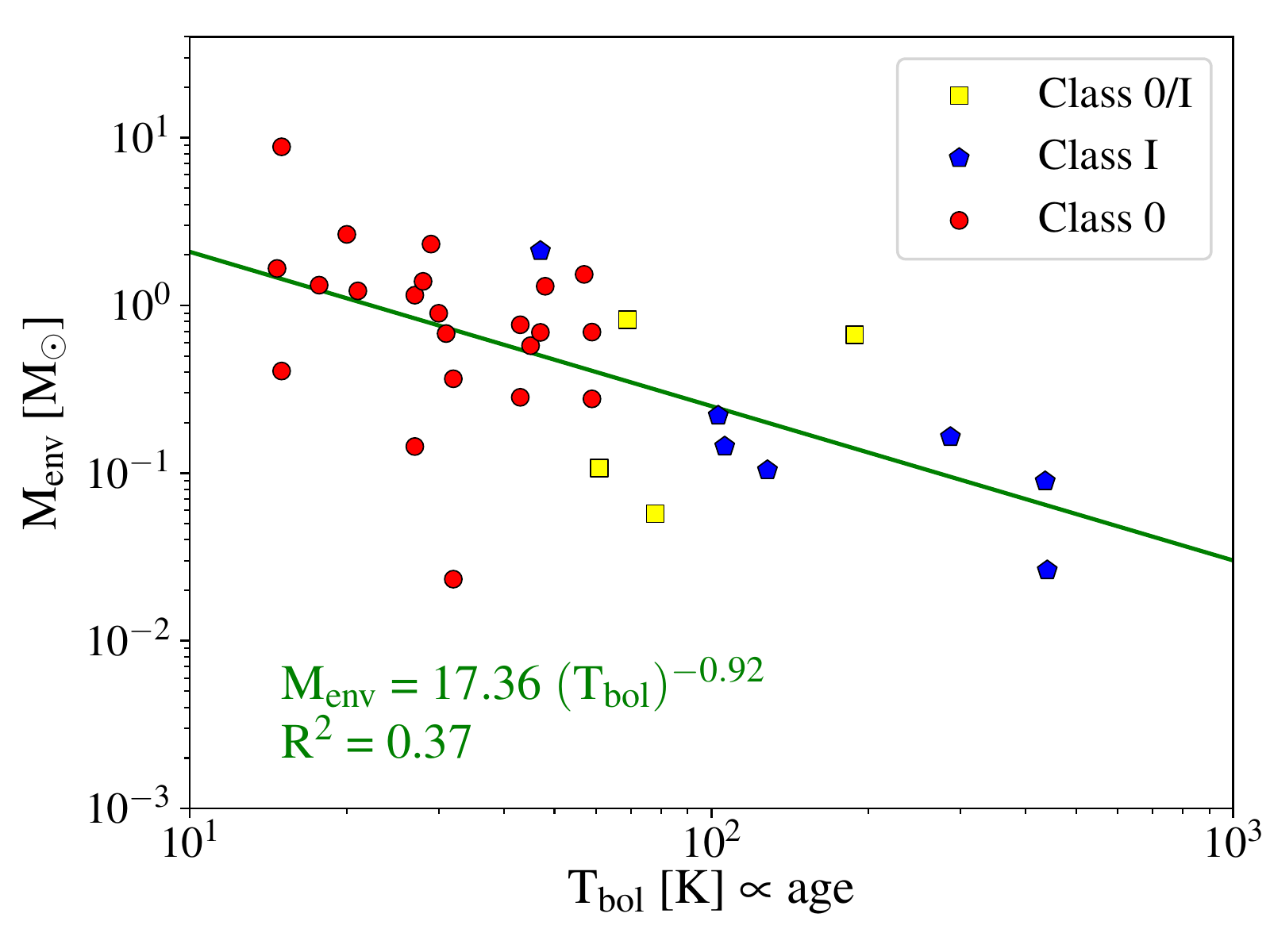}
    \caption{Estimated envelope masses as a function of bolometric temperature, with the colors of the points corresponding to protostellar classification (see Figure~\ref{sc_compare}). The masses exhibit a negative trend ($R^2$~$=$~$0.37$) described by the power-law line plotted in green. This negative trend represents envelope dissipation over time.} \label{m_env_Tbol}
\end{figure}

\begin{figure}[!ht]
  	\centering
    \includegraphics[width=0.5\textwidth]{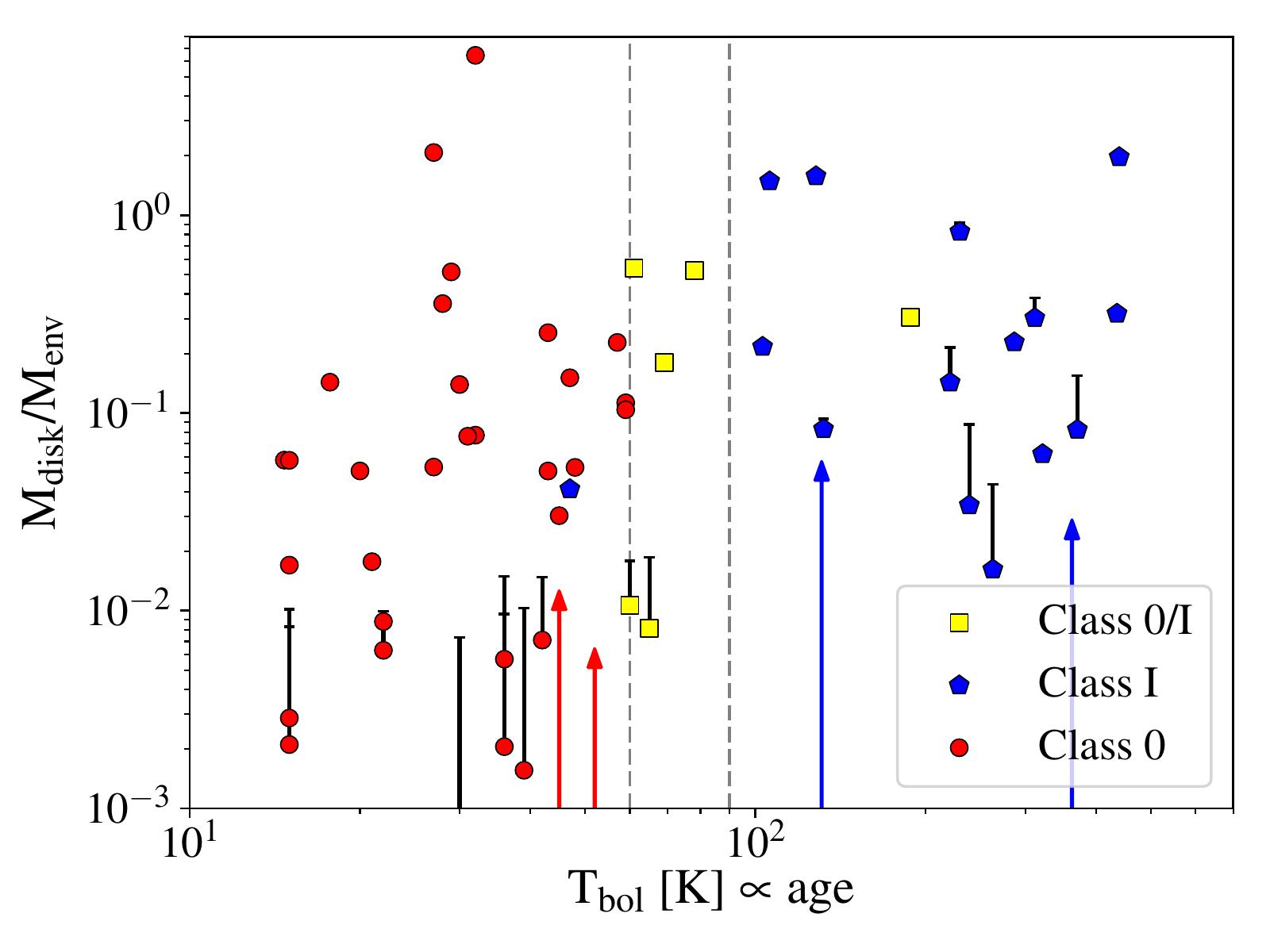}
    \caption{Ratio of the disk mass to the envelope mass as a function of bolometric temperature, with the colors of the points corresponding to protostellar classification (see Figure~\ref{sc_compare}). The data exhibits a weak positive trend ($R^2$~$=$~$0.21$) corresponding to envelope dissipation. Each error bar represents the upper limit mass ratio measurement for a given source, calculated using the disk mass upper bound and corresponding envelope mass. See the description under Figure \ref{m_disk_Tbol} for more details.} \label{m_ratio_Tbol}
\end{figure}

\indent Figure \ref{m_disk_Tbol} shows the estimated disk mass as a function of bolometric temperature for each of the $59$ sources in our sample. The disk mass upper bounds described in Section 3.4 are displayed as error bars extending from the measured disk mass to the $3\sigma_{M_{\text{disk}}}$ upper bound value. The bolometric temperatures used in this plot are taken from \cite{Tobin2016} and references therein. \cite{Tobin2016} labels sources within $60$\,K $ \leq \text{T}_{\text{bol}} \leq 90$\,K as Class 0/I since measured bolometric temperature has an uncertainty dependent on inclination angle that makes classification within this temperature range ambiguous. In Figure \ref{m_disk_Tbol}, Class 0/I sources are labeled as yellow squares. The vertical dashed gray lines separate Class 0, Class 0/I, and Class I sources based solely on bolometric temperature cutoffs. Note that there appear to be two sources that are not properly sorted based on these vertical lines. \cite{Tobin2007,Tobin2016} determined these two source classifications (for Per-emb-44 and L1448IRS3A) based on criteria other than bolometric temperature. \newline
\indent The derived disk masses for all protostellar classifications range from ${\sim}\,0.001$ to $1.2$\,M$_{\odot}$. Including the sources with upper limits, the median Class 0 disk mass is $0.04$\,M$_{\odot}$ and the median Class I disk mass is slightly lower at $0.03$\,M$_{\odot}$. While these typical masses are roughly in agreement with those found by J09, \cite{Enoch2011}, and F17, our data reveal a subgroup of Class 0 and Class 0/I sources at low disk masses that has not been explicitly identified in previous studies. This subgroup is separated from all other sources in the sample, bound by bolometric temperatures less than $100$\,K and upper limit disk mass estimates of less than ${\sim}\,0.008$\,M$_{\odot}$. All but one of these sources (Per-emb-37) is a single system (Tobin et al. 2016), suggesting that we probably do not over-estimate (and thus over-subtract) the single-dish flux due to confusion with other protostars. None of these particularly low-mass sources were identified by S18 as a disk candidate, indicating that their disks, if they exist, may be too low-mass or compact to be fit by disk models in the VANDAM survey. This subgroup may indicate that some young sources have their early disk production inhibited by some physical phenomenon, leading to relatively small, low-mass disks. Such physical phenomena may include magnetic braking \citep[e.g.,][]{Allen2003}, non-ideal MHD effects, or one or more past encounters with other objects.
\newline
\indent It is worth noting that a good portion of Class I sources are below their $3\sigma_{M_{\text{disk}}}$ significant detection upper bounds. Excluding all sources with disk mass upper limit measurements, a least-squares power-law fit to the rest of the data in Figure~\ref{m_disk_Tbol} reveals no significant trend, with a coefficient of determination of only $R^2 = 0.001$. This is consistent with the results from J09 and \cite{Enoch2011}, in which they found no evidence for significant mass growth between the Class 0 and Class I sources in their samples. This lack of a trend combined with the separation of the low-mass subgroup may indicate the existence of two distinct modes of early disk formation: 1) a model in which disk growth in collapsing protostellar systems is not significantly inhibited by some phenomenon, leading to rapid disk formation in the \textit{early} Class 0 embedded stage with disk masses remaining roughly constant over time and 2) a model in which disk growth in collapsing protostellar systems is significantly inhibited, leading to the formation of the low-mass subgroup of disks. \newline
\indent Figure \ref{m_env_Tbol} compares the estimated envelope mass as a function of bolometric temperature, for each of the $33$ sources in our sample without disk mass upper bounds and with nonnegative envelope masses. We find a range of envelope masses extending from ${\sim}\,0.02$ to $8.8$\,M$_{\odot}$. The median Class 0 envelope mass is $0.73$\,M$_{\odot}$ while the median Class I envelope mass is significantly lower at $0.10$\,M$_{\odot}$. Qualitatively, the plot reveals a negative trend in the envelope mass going from Class 0 to Class I sources. By fitting a least-squares power-law to the data, we confirm this trend with a coefficient of determination of $R^2$~$=$~$0.37$. The resulting fit power-law is shown in Figure \ref{m_env_Tbol} as a green line. This negative trend is likely a manifestation of envelope dissipation over time as matter is accreted from the envelope onto the disk and into the central star (e.g., J09) and possibly dispersed by outflows \citep[e.g.,][]{Arce2006,Offner2014}. \newline
\indent Figure \ref{m_ratio_Tbol} shows the ratio of the disk mass divided by the envelope mass versus bolometric temperature. As expected from the previous plots, since disk mass remains relatively constant with respect to $\text{T}_{\text{bol}}$ while envelope mass tends to decrease, then the ratio $M_{\text{disk}}/M_{\text{env}}$ tends to slightly increase over time. The coefficient of determination of a least-squares power-law fit to the data (excluding all sources with mass ratio upper limits) is only $R^2$~$=$~$0.21$, indicating a weak correlation. This relation reflects the fact that the envelopes are dissipating over time.

\section{Summary} \label{summary}
\indent In this paper we used the J09 method to separate the disk and envelope masses for $59$ Class 0 and Class I sources in the Perseus molecular cloud. We extracted compact and extended fluxes for each source using SMA and JCMT/SCUBA-2 data, respectively. Then we used the system of equations proposed by J09 to isolate the disk and envelope fluxes. Assuming optically thin disk emission, we converted these isolated fluxes into independent disk and envelope mass estimates. We then compared our results to independent disk mass estimates from the resolved continuum observations in the VANDAM survey and looked for trends in our own data. Our primary results are the following:
\begin{enumerate}
\item A linear least-squares fit of our disk masses compared to disk masses estimated from the VANDAM survey yields a strong linear correlation ($R^2$~$=$~$0.97$ for the Segura-Cox masses and $R^2$~$=$~$0.92$ for the Tychoniec masses). The strength of the linear correlation between our results suggests that the J09 method is likely valid for measuring disk masses with unresolved continuum observations at ${\sim}\,1000$\,AU resolution.
\item We detect sources with disk masses ${>}0.1$\,M$_{\odot}$ that were not identified by S18 as disk candidates. In particular, our estimated disk mass for Per-emb-13 is ${\sim}\,0.5$\,M$_{\odot}$. For some of these sources it is possible that large dust grains in the disk have become condensed towards the center, so that the disk appears unresolved in the $8$\,mm VANDAM survey. Other sources are certainly morphologically complex and difficult to fit with a simple disk.
\item Our comparisons of disk and envelope masses versus bolometric temperature show that disk masses remain roughly constant between Class 0 and Class I sources while the envelope mass dissipates over time. In addition, we identify a possible group of young, low-mass ($\text{T}_{\text{bol}}$~$<$~$100$\,K and $M_{\text{disk}}$~$\lesssim$~$0.008$\,M$_{\odot}$) sources separated from the rest of the population. These results hint at the existence of two distinct modes of early disk formation: 
\begin{enumerate}
    \item A mode where disk growth in collapsing protostellar systems is not significantly inhibited by some physical phenomenon (for example, magnetic braking), suggesting disk formation occurs rapidly during the \textit{early} Class 0 embedded stage. After formation, disk masses remain roughly constant over time.
    \item A mode where disk growth in collapsing protostellar systems is significantly inhibited in the early Class 0 stage, leading to the formation of the observed low-mass subgroup of disks.
\end{enumerate}
\end{enumerate}
\indent Our analysis supports the validity of the J09 method. In the future, this method can be applied to measure disk masses of protostars in other molecular clouds using unresolved data.

\acknowledgments
\noindent \textit{Acknowledgements:} This research was conducted under the Smithsonian Astrophysical Institute (SAO) Research Experience for Undergraduates (REU) program. The SAO REU program is funded by the National Science Foundation REU and Department of Defense ASSURE programs under NSF Grant AST-1659473, and by the Smithsonian Institution.

Astrochemistry in Leiden is supported by the Netherlands Research School for Astronomy (NOVA), by a Royal Netherlands Academy of Arts and Sciences (KNAW) professor prize, and by the European Union A-ERC grant 291141 CHEMPLAN.

The Submillimeter Array is a joint project between the Smithsonian Astrophysical Observatory and the Academia Sinica Institute of Astronomy and Astrophysics, and is funded by the Smithsonian Institution and the Academia Sinica. We also wish to recognize and acknowledge the very significant cultural role and reverence that the summit of Maunakea has always had within the indigenous Hawaiian community. We are most fortunate to have the opportunity to conduct observations from this mountain.

We would also like to acknowledge the technical support provided by Mark Gurwell, Charlie Qi, and Glen Petitpas. 

The JCMT has historically been operated by the Joint Astronomy Centre on behalf of the Science and Technology Facilities Council of the United Kingdom, the National Research Council of Canada and the Netherlands Organisation for Scientific Research. Additional funds for the construction of SCUBA-2 were provided by the Canada Foundation for Innovation. We thank the JCMT staff for their support of the GBS team in data collection and reduction efforts. We also thank Sarah I. Sadavoy and Helen Kirk for sharing data before their release date and for discussing the details of the data.

This research used the services of the Canadian Advanced Network for Astronomy Research (CANFAR) which in turn is supported by CANARIE, Compute Canada, University of Victoria, the National Research Council of Canada, and the Canadian Space Agency. We also used the facilities of the Canadian Astronomy Data Centre operated by the National Research Council of Canada with the support of the Canadian Space Agency.



\end{document}